\documentclass[a4paper,12pt]{article}
\usepackage[utf8]{in putenc}
\usepackage{cancel}
\usepackage{ulem}
\usepackage{amsfonts}
\usepackage{amssymb}
\usepackage{graphicx}
\usepackage{amsmath}
\usepackage{enumerate}
\usepackage{mathtools}
\usepackage{pgfplots}
\usepackage{subfig}
\usepackage{color}
\usepackage{float}
\usepackage{here}
\usepackage{cite}
\usepackage{mathrsfs}
\usepackage{float,epsfig}
\usepackage{dcolumn}
\usepackage{tikz}\usetikzlibrary{calc}
\usetikzlibrary{decorations.pathreplacing}
\usepackage{graphicx}
\usepackage{bm}
\usepackage{amsmath,amssymb,amsthm}
\usepackage[colorlinks=true,linkcolor=blue,citecolor=red]{hyperref}
\newcommand{\be}{\begin{equation}}
\newcommand{\ee}{\end{equation}}
\newcommand{\bea}{\setlength\arraycolsep{2pt} \begin{eqnarray}}
\newcommand{\eea}{\end{eqnarray}}

\def\0{{\sst{(0)}}}
\def\1{{\sst{(1)}}}
\def\2{{\sst{(2)}}}
\def\3{{\sst{(3)}}}
\def\4{{\sst{(4)}}}
\def\5{{\sst{(5)}}}
\def\6{{\sst{(6)}}}
\def\7{{\sst{(7)}}}
\def\8{{\sst{(8)}}}
\def\sst#1{{\scriptscriptstyle #1}}

\title{\bf \Large Thermal Image 
and Phase Transitions of Charged AdS Black Holes  using  Shadow Analysis}

\author{
A.  Belhaj$^{1}$\thanks{belhajadil@fsr.ac.ma}, L. Chakhchi$^{1}$\thanks{chakhchi.07@gmail.com}, H.  El Moumni$^{2}$\thanks{hasan.elmoumni@edu.uca.ma (Corresponding author)},
J. Khalloufi$^3$\footnote{jamalkhalloufi@gmail.com}
,  K.  Masmar$^4$\footnote{karima.masmar@edu.uca.ac.ma},
 \footnote{ Authors in alphabetical order, they contributed equally to this work.}
\\
{\small $^{1}$ ESMAR, Physics Department, Faculty of Science,
Mohammed $V$ University in Rabat, Morocco. }\\
{\small $^{2}$ EPTHE, Physics Department, Faculty of Science,  Ibn Zohr University, Agadir, Morocco. }\\
{\small $^{3}$ LPHEA, FS,
	Cadi Ayyad University,  Marrakech, Morocco.}\\
  {\small $^{4}$ Laboratory of  High Energy Physics and Condensed Matter
Hassan II University} \\{\small Faculty of Science Ain Chock, Casablanca, Morocco. }
}

\vspace{-3.6em}

\begin{document}
\maketitle
\begin{abstract}
We investigate the relations between the black hole shadow and charged AdS black hole critical behavior in the extended phase space.  Using  the  thermo-shadow formalism  built in \cite{Zhang:2019glo}, we reveal that the shadow radius can be considered as an efficient tool to study thermodynamical black hole systems. Based on such arguments,  we build a thermal profile by varying the RN-AdS black hole temperature on the shadow silhouette.  Among others, the Van der Waals-like phase transition takes place.
This could open a new window on the thermal picture of black holes and the corresponding thermodynamics from the observational point of view.

{\noindent}

\end{abstract}
\tableofcontents


\section{Introduction}

The Event Horizon Telescope (EHT) image has been considered as a relevant achievement of the Einstein theory of gravitation\cite{Akiyama:2019cqa,Akiyama:2019eap,Akiyama:2019fyp}. The  black hole image,  in the center of  M$87^\star$ galaxy,  provides huge and immense supports  to its existence in the cosmos parallelly with gravitational wave detections by LIGO and Virgo \cite{Abbott:2016blz}.  It has opened a new window to understand the elements of the gravitation theory and brought  primordial information concerning jets and matter dynamics around such compact objects.  Precisely, the shadow of a black hole can be interpreted as an abundant source of data describing the black hole hairs  including the mass, the charge, and the rotation physical parameters. This can be exploited to unveil physical constraints on the gravitational theories \cite{Cunha:2019ikd,Vagnozzi:2019apd,Allahyari:2019jqz}.  More precisely, the black hole shadow  corresponds to  the gravitational lensing  of light around massive objects. Interposing the black hole between an observer and   certain sources,  it has been observed that not all emitted  lights    can come to  the observer after communication with  the strong gravitational field of the black hole.  In this way,  certain  photons collapse into the black hole generating the so  called  shadow.   Many investigations have been elaborated including the seminal works
\cite{synge1966escape,luminet1979image}
dealing with
the shadow of a Schwarzschild black hole  and   extensions to  rotating   cases \cite{synge1966escape,luminet1979image,hawking1973black,de2000apparent,Hioki:2008zw,Li:2020drn,Guo:2019lur,Shaikh:2018lcc}.  In particular, this direction has  been extensively investigated  using different ways and methods \cite{Grenzebach:2014fha,Haroon:2018ryd,Johannsen:2015hib,Cunha:2015yba,Eiroa:2017uuq,Wang:2017qhh,
Tsukamoto:2017fxq,Tsupko:2017rdo,Sharif:2016znp,Ohgami:2016iqm,Younsi:2016azx,Abdujabbarov:2016hnw,
Amir:2016cen,Atamurotov:2015nra,Perlick:2015vta,Moffat:2015kva,Lu:2014zja,Atamurotov:2013sca,Guo:2018kis,
Yan:2019etp,Hennigar:2018hza,Konoplya:2019sns,Bambi:2008jg,Bambi:2010hf,Konoplya:2019fpy,Bambi:2019tjh}.


Recently,  it has been remarked that  the study of   thermodynamic proprieties of  black holes,  in  AdS spacetime,  has received  remarkable  interests \cite{Chamblin:1999tk,Kubiznak:2012wp,Belhaj:2012bg,Gunasekaran:2012dq,Belhaj:2015hha,Hendi:2012um,Zhang:2015ova,Wei:2015iwa,Chabab:2015ytz, Perlick:2018iye,Nguyen:2015wfa,moiplb,Liu:2014gvf,Chabab:2016cem,Zou:2017juz,Chabab:2018lzf, Wei:2017mwc,Chabab:2019kfs
}. Interpreting the cosmological constant in such a space  as the pressure of the general thermodynamic system $P=-\frac{\Lambda}{8\pi}=\frac{3\pi}{8\ell^2}$,  many interesting results have been obtained  for various   black hole solutions.   Concretely,  it has been found a nice interplay between the black hole state parameter and the Van der Waal's  state equation  generating a  critical phenomenon in AdS spacetime backgrounds.

 More recently, such issues   have been linked to shadow formalism. In particular,  it has been established a direct bridge between the  black hole thermodynamical systems  and the shadow black holes \cite{Zhang:2019glo}.

The aim of this work is to contribute to these activities by investigating the thermal profile of AdS black hole shadows.  Precisely, we study the correlation between the black hole shadows and RN- AdS black hole critical behaviors  in the extended phase space.   Based on the associated  developed analysis, we show that the shadow radius can be considered as an efficient tool to study thermodynamical black hole systems.  In particular,   we construct a thermal profile by varying the RN-AdS black hole temperature on the shadow silhouette, where the Van der Waals-like phase transition takes place.
We expect that this could open a  new gate for unveiling the thermal picture of black holes and the corresponding thermodynamics from the observational point of view.

The paper is structured as follows.  In section.\ref{satp},  we give a concise  review on shadow formalism applied to AdS black holes.  In   section.\ref{sec3},  we  show that the Van der Waals-like phase transition in $T-r_h$
 persists in $T-r_s$ plane. In section.\ref{sec4}, we present a  thermal profile of the shadow associated with the RN-AdS black hole.  In particular, we show   that the oscillation behavior of the temperature characterizing the Van der Waals-like transition can be directly observed in such a  thermal picture. The last section is  reserved  to  conclusions and certain open questions.

\section{Shadows and thermodynamics of black holes}
\label{satp}
\begin{figure}[!ht]
 \center
  \includegraphics[scale=.65]{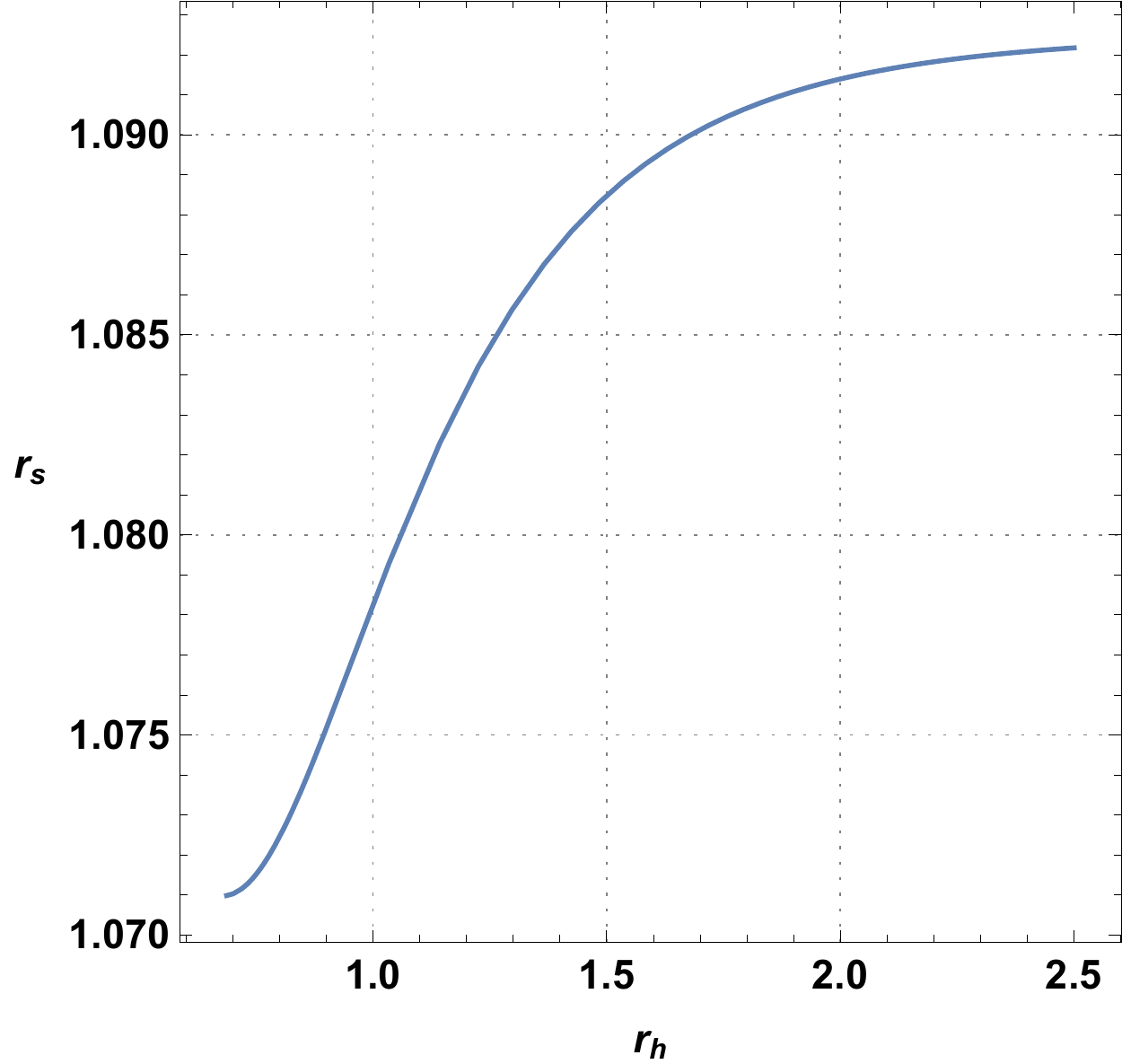}\\
  \caption {\it{ \footnotesize
The variation of  shadow radius $r_s$ in  in terms of the event horizon radius $r_h$, here we have set $Q=1$ and $P=0.1$.}}
\label{fig2}
\end{figure}

In this section, we give a concise review on  the known results on  the black hole shadow and the corresponding  thermodynamics being firstly  developed in \cite{Zhang:2019glo}.  For  more details, we refer to such a work. Indeed, we consider   a static spherically symmetric space-time background which can be described by the following  line element expression
\begin{equation}
ds^{2}=-f(r)dt^{2}+\frac{dr^{2}}{g(r)}+r^{2}d\theta^{2}+r^{2}\sin^{2}\theta d\phi^{2}.
\end{equation}
$f(r)$ and $g(r)$ stand for the blacking functions which depend on the variable  $r$.  Following \cite{Zhang:2019glo,Perlick:2018iye},  the  Hamiltonian describing   a massless  photon moving reads as 
\begin{equation}\label{hamij}
\mathcal{H}=\frac{1}{2}g^{ij}p_{i}p_{j}=0.
\end{equation}
Considering a  photon motion on the  equatorial plane required by $\theta=\pi/2$,  Eq.\eqref{hamij}  reduces to
\begin{equation}\label{ramo}
\frac{1}{2}\left[-\frac{p_{t}^{2}}{f(r)}+g(r)p_{r}^{2}+\frac{p^{2}_{\phi}}{r^{2}}\right]=0.
\end{equation}
Since  Hamiltonian does not depend explicitly on the coordinates $t$ and $\phi$,  the  associated  two constants of motion   can be written as 
\begin{equation}
p_{t}\equiv\frac{\partial \mathcal{H}}{\partial \dot{t}}=-\mathcal{E}\quad
\text{and}\quad
p_{\phi}\equiv\frac{\partial \mathcal{H}}{\partial \dot{\phi}}=\mathcal{J}.
\end{equation}
In this way,  the quantities $\mathcal{E}$ and $\mathcal{J}$  are considered as the energy and  the angular momentum of the photon,  respectively. Exploiting the Hamiltonian formalism, the  associated  equations of motion are 
\begin{equation}
\dot{t}=\frac{\partial H}{\partial p_{t}}=-\frac{p_{t}}{f(r)},
\quad
\dot{\phi}=\frac{\partial H}{\partial p_{\phi}}=\frac{p_{\phi}}{r^{2}},
\quad \text{and}\quad
\dot{r}=\frac{\partial H}{\partial p_{r}}=p_{r}g(r),
\end{equation}
Here,  the  over dot is the derivative with respect to  the affine parameter $\tau$ and   $p_{r}$  is   the radial momentum.  They  provide  a complete description of the dynamics where  the  effective potential  reads as
\begin{equation}\label{e1}
V_{e}(r)+\dot{r}^{2}=0,
\end{equation}
from which one obtains
\begin{equation}\label{e1}
 V_{e}(r)=g(r)\left[\frac{\mathcal{J}^{2}}{r^{2}}-\frac{\mathcal{E}^{2}}{f(r)}\right].
\end{equation}
It is worth nothing that  the  photon spherical  geometry  is  associated with   the constraints
\begin{equation}\label{effpho}
V_{e}(r_p)=0,\quad  \left.\frac{\partial V_{e}(r)}{\partial r}\right|_{r=r_p}=0\quad \text{and}\quad  \left.\frac{\partial^2 V_{e}(r)}{\partial r^2}\right|_{r=r_p}>0.
\end{equation}
It has been shown that this   produces an equation  parameterizing  the  orbit of the photon
\begin{equation}\label{oreq}
\frac{dr}{d\phi}=\frac{\dot{r}}{\dot{\phi}}=\frac{r^{2}g(r)p_{r}}{\mathcal{J}}.
\end{equation}
Using  Eq.\eqref{ramo},  an  explicit form for   the Eq.\eqref{oreq} can be obtained  as follows
\begin{equation}
\frac{dr}{d\phi}=\pm r\sqrt{g(r)\left[\frac{r^{2}\mathcal{E}^{2}}{f(r)\mathcal{J}^{2}}-1\right]}.
\end{equation}
The turning point of the photon orbit being  interpreted mathematically by the constraint $\left.\frac{dr}{d\phi}\right|_{r=R}=0$  gives 
\begin{equation}\label{eosf}
\frac{dr}{d\phi}=\pm r\sqrt{g(r)\left[\frac{r^{2}f(R)}{f(r)R^{2}}-1\right]}.
\end{equation}
For later use, we consider  a light ray sending from a static observer placed  at $r_{o}$ and transmitting into the past with an angle $\alpha$ with respect to the radial direction. 
In this way,  we have
\begin{equation}
\cot\alpha=\frac{\sqrt{g_{rr}}}{\sqrt{g_{\phi\phi}}}\frac{dr}{d\phi}{\Big{|}}_{r=r_{o}}=\frac{1}{r\sqrt{g(r)}}\frac{dr}{d\phi}{\Big{|}}_{r=r_{o}}.
\end{equation}
Exploiting Eq.\eqref{eosf},    one obtains 
\begin{equation}
\cot^{2}\alpha=\frac{r_{o}^{2}f(R)}{f(r_{o})R^{2}}-1,  \quad  \sin^{2}\alpha=\frac{f(r_{o})R^{2}}{r_{o}^{2}f(R)}.
\end{equation}
In this context,  one gets  the  angular radius of the black hole shadow  by sending   $R$ to $r_{p}$  being the circular orbit radius of the photon appearing in   Eq.\eqref{effpho}. Precisely,  the  shadow radius of the black hole observed by a static observer placed  at $r_{o}$  is given by
\begin{equation}\label{shara}
r_{s}=r_{o}\sin\alpha=\left. R\sqrt{\frac{f(r_{o})}{f(R)}}\right|_{R=r_{p}}.
\end{equation}
According to  \cite{Zhang:2019glo},  a  link between the black hole shadows and the black hole thermodynamics has been established. Exploiting  the heat capacity
\begin{equation}\label{C}
C=T\left(\frac{\partial S}{\partial T}\right)
\end{equation}
and using the fact that  the entropy  is related to event horizon by  $S=\pi r_h^2$, with $\frac{\partial S}{\partial r_h}>0$, the sign of $C$ is  directly  deduced  from the one  of $\frac{\partial T}{\partial r_h}$.  The computation leads to 
\begin{equation}\label{variation}
\frac{d T}{d r_h}= \frac{d T}{d r_s}      \frac{d r_s}{dr_h}.
\end{equation}
When the  constraint $\frac{d r_s}{dr_h}>0$ is  satisfied, one can consider
\begin{equation}
\text{Sgn}(C)= \text{Sgn}\left(\frac{\partial T}{\partial r_s}\right).
\end{equation}
Having discussed the associated  shadow backgrounds,  we move  to prob the link  between the shadow radius of the black hole and its phase transition structure including  the Van der Waals like phase one.  To do so, we consider  an AdS  black hole solution with a charge $Q$  corresponding to the  following   blackening function
\begin{equation}\label{met1}
f(r)=g(r)=1-\frac{2M}{r}+\frac{Q^{2}}{r^{2}}+\frac{8 \pi P r^2}{3},
\end{equation}
where $M$ is   the corresponding  mass.  It is noted that $P$ is linked  to  the  AdS radius $\ell$ via $P=\frac{3}{8\pi\ell^2}$ \cite{Kubiznak:2012wp,Gunasekaran:2012dq}. The event horizon   is associated with  the large root of the equation $f(r)|_{r=r_h}=0$, while the Hawking temperature is obtained from the relation
\begin{equation}\label{temp}
T=\left.\frac{f'(r)}{4\pi}\right|_{r=r_h}=\frac{1}{4 \pi  r_h}-\frac{Q^2}{4 \pi  r_h^3}+2 P r_h.
\end{equation}
The heat capacity can be expressed as
\begin{equation}\label{CC}
C=\frac{2 \pi  r_h^2 \left(8 \pi  P r_h^4-Q^2+r_h^2\right)}{8 \pi  P r_h^4+3 Q^2-r_h^2}.
\end{equation}
Using the  constraint on the effective potential Eq.\eqref{effpho}, we can obtain   the  formula of  the photon  circular orbit radius
\begin{equation}\label{phrnx}
r_{p}=\frac{1}{2} \left(3M+\sqrt{9 M^2-8 Q^2}\right).
\end{equation}
It follows  that this does not depend on  the AdS radius $\ell$. Eq.\eqref{shara} allows one to get  the RN-AdS black hole shadow as follows
\begin{eqnarray}\label{rs}\nonumber
r_{s}&=&\frac{\sqrt{6 f(r_o)} Q \left(\sqrt{9 M^2-8 Q^2}+3 M\right)}{\sqrt{9 M^2 \left(32 \pi  P
   Q^2-1\right)+3 M \sqrt{9 M^2-8 Q^2} \left(32 \pi  P Q^2+1\right)+4 Q^2 \left(3-32 \pi
   P Q^2\right)}}.\\
\end{eqnarray}
An observer at spatial infinity  can be mathematically translated  by imposing the constraint $f(r_0)=1$, giving the following form of the shadow radius
\begin{eqnarray}\label{rs}\nonumber
r_{s}
&=&\frac{\sqrt{6} Q \left(\sqrt{9 M^2-8 Q^2}+3 M\right)}{\sqrt{9 M^2 \left(32 \pi  P
   Q^2-1\right)+3 M \sqrt{9 M^2-8 Q^2} \left(32 \pi  P Q^2+1\right)+4 Q^2 \left(3-32 \pi
   P Q^2\right)}}\\
\end{eqnarray}
In the vanishing limit of pressure/cosmological constant  $P\rightarrow 0$, associated with a large AdS  spacetime radius $\ell\rightarrow\infty$, we recover the result  reported in  \cite{Zhang:2019glo}. To cheek  the condition $\frac{dr_s}{dr_h}>0$  from   Eq.\eqref{variation}, we  illustrate  the variation of the shadow radius $r_s$ in terms of the horizon radius $r_h$ in Fig.\ref{fig2}.
It follows from this figure  that $r_s$  increases   in terms of    $r_h$ ensuring that the positivity of the first derivative of $r_s$ with respect of  $r_h$.  Eq.\eqref{variation} shows that  the   temperature exhibits similar behaviors  for both variables   $r_s$ and  $r_h$.

The rest of paper is the main  investigation part. In particular, we study     the phase transition of AdS black holes using  shadow analysis.

\section{Phase  transition of charged AdS black hole using  shadow formalism}\label{sec3}

\begin{figure}[!ht]
		\begin{center}
		\centering
			\begin{tabbing}
			\centering
			\hspace{9.3cm}\=\kill
			\includegraphics[scale=.5]{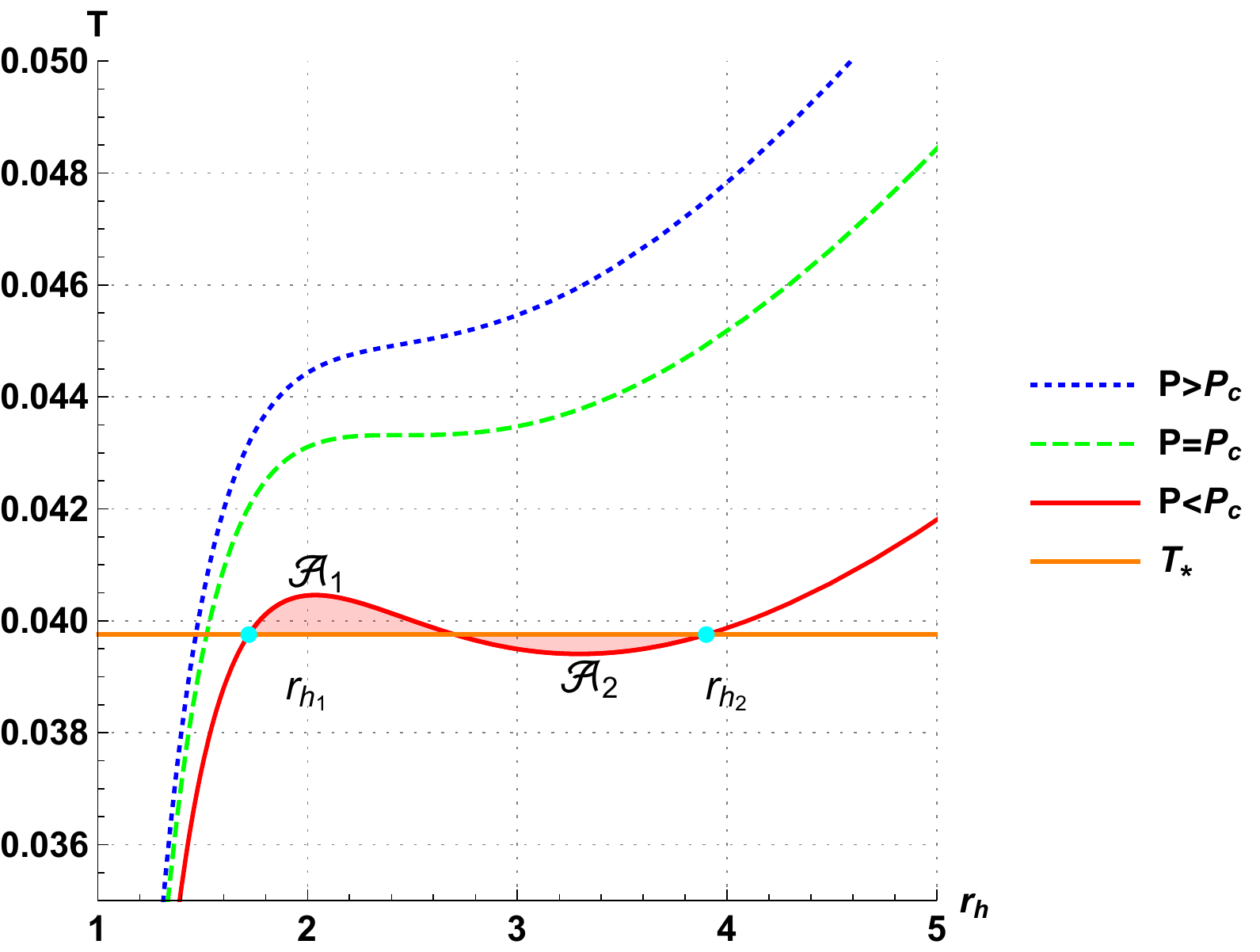} \>
			\includegraphics[scale=.5]{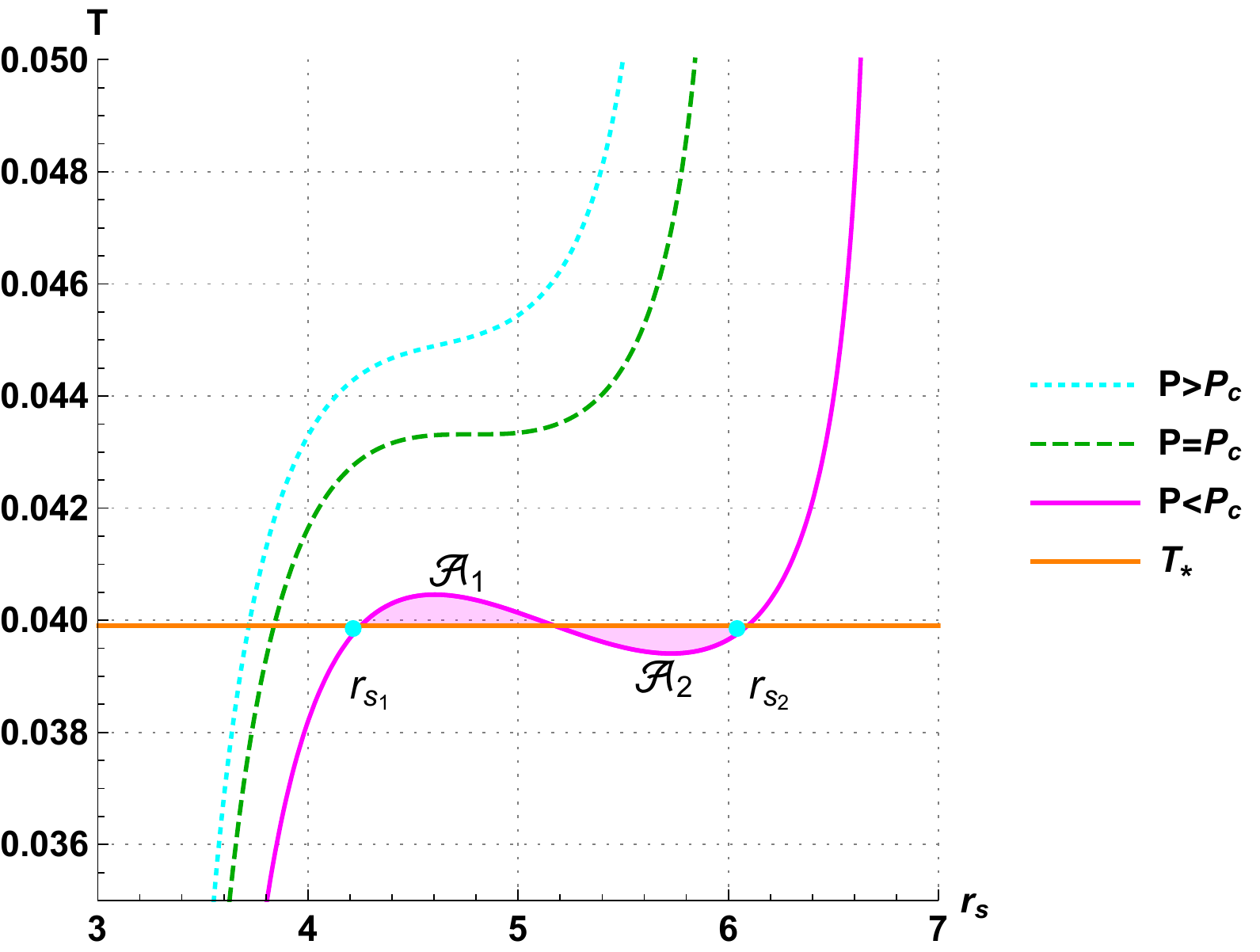} \\
			\includegraphics[scale=.5]{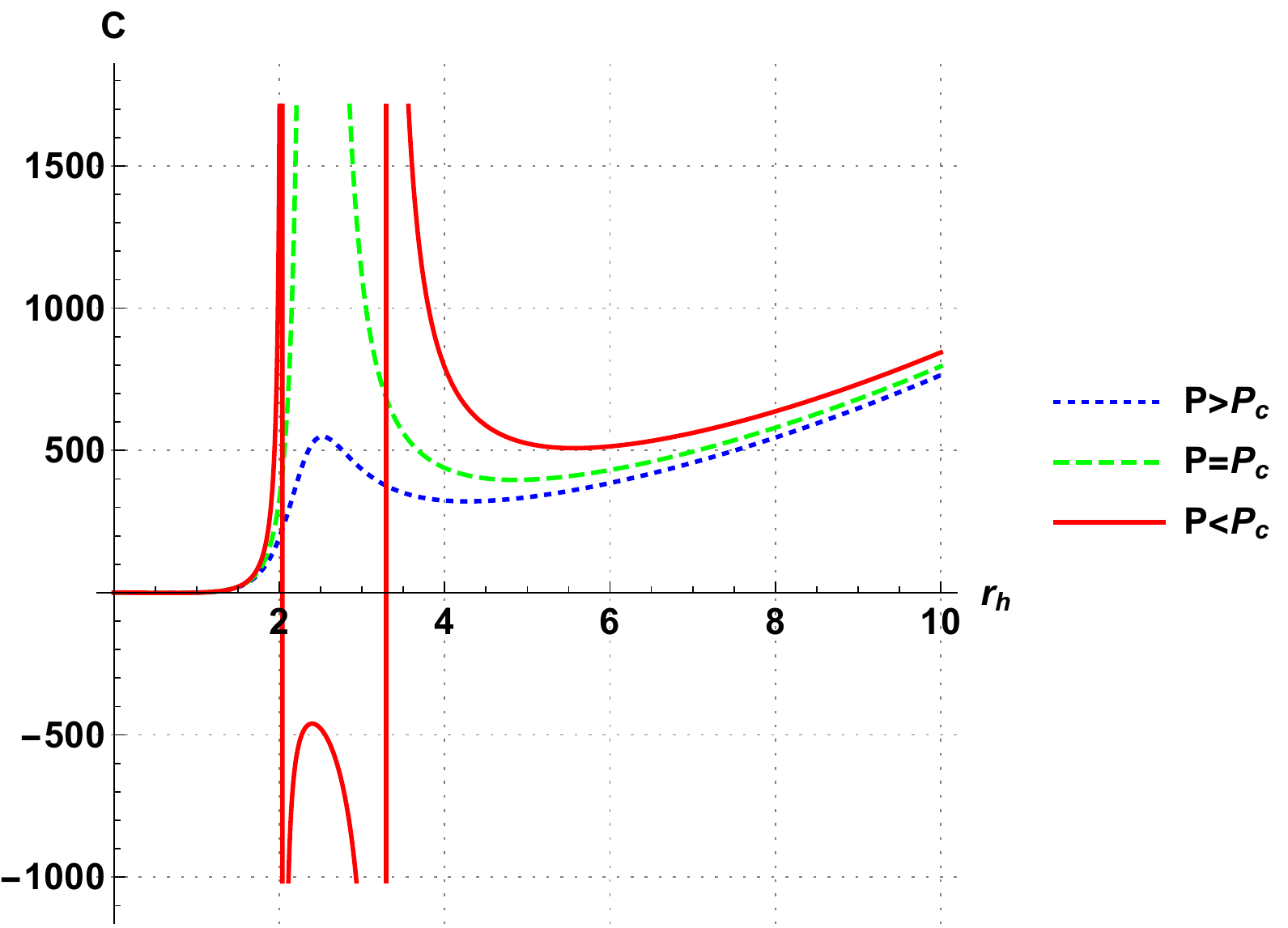} \>
			\includegraphics[scale=.5]{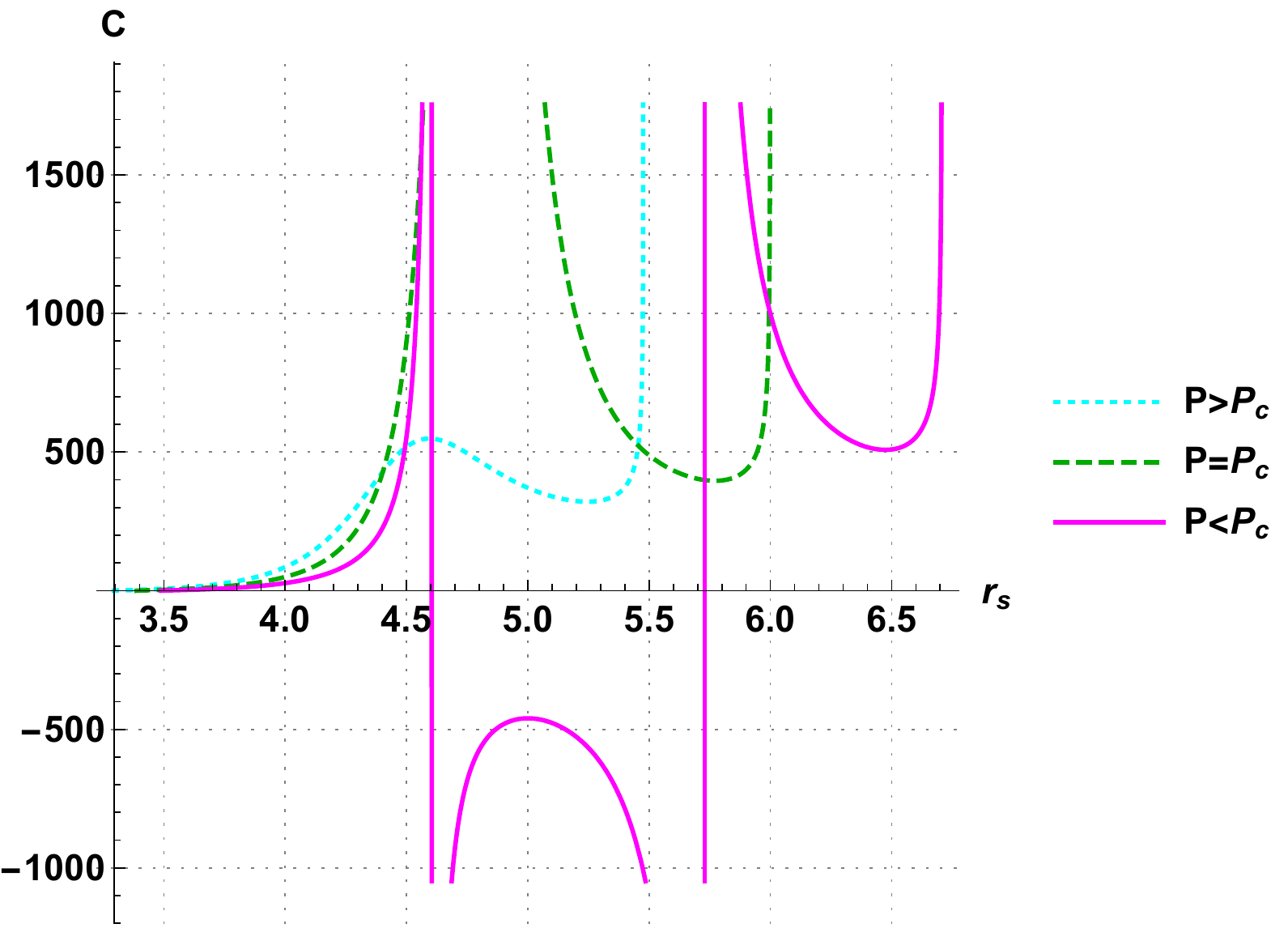} \\
		   \end{tabbing}
\caption{{\it \footnotesize {\bf Top:} The relation between horizon radius $r_h$ ({\bf left}), shadow radius $r_s$ ({\bf right}) and temperature for different values of the pressure. The solid line corresponds to the location of the first-order phase transition, while the dashed line to the second-order one. $T_\star$ is the coexistence temperature.
 {\bf Buttom:} The heat capacity versus $r_h$ ({\bf left}) and $r_s$ ({\bf right})  for different values of the pressure. For all panels, we have set the charge $Q=1$.   }}\label{Tradii}
\end{center}
\end{figure}

\begin{figure}[!ht]
		\begin{center}
		\begin{tikzpicture}[scale=0.2,text centered]
\node[] at (-20,1){\small  \includegraphics[scale=0.5]{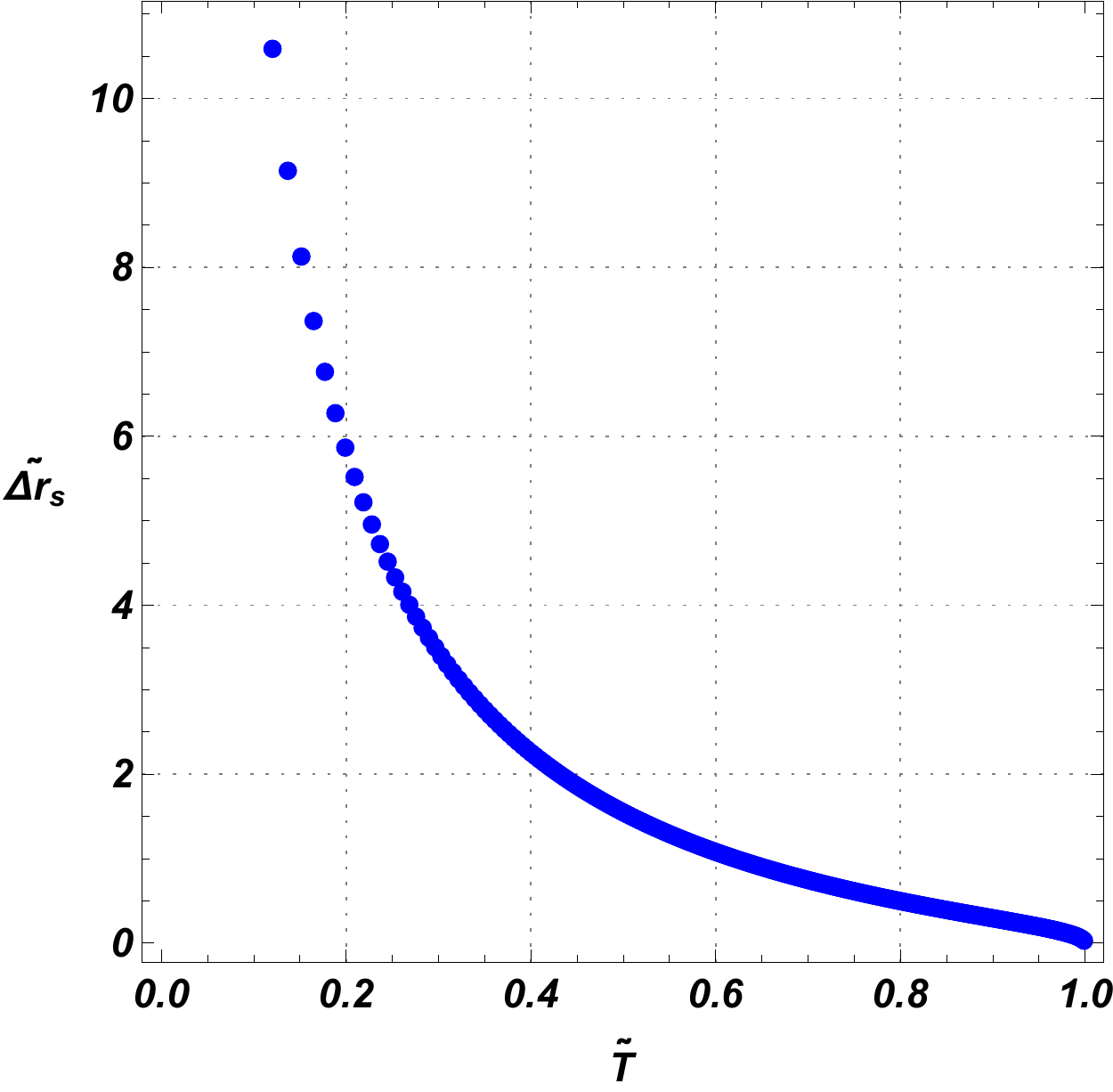}};	

\node[draw,circle ,thin,line width=1pt,color=magenta,minimum size=.65cm] at (-5.,-9.8){ \textcolor{black}{\tiny }};

\node[draw, thin,line width=0pt,color=magenta,name=plan] at (24,1){\small  \includegraphics[scale=0.5]{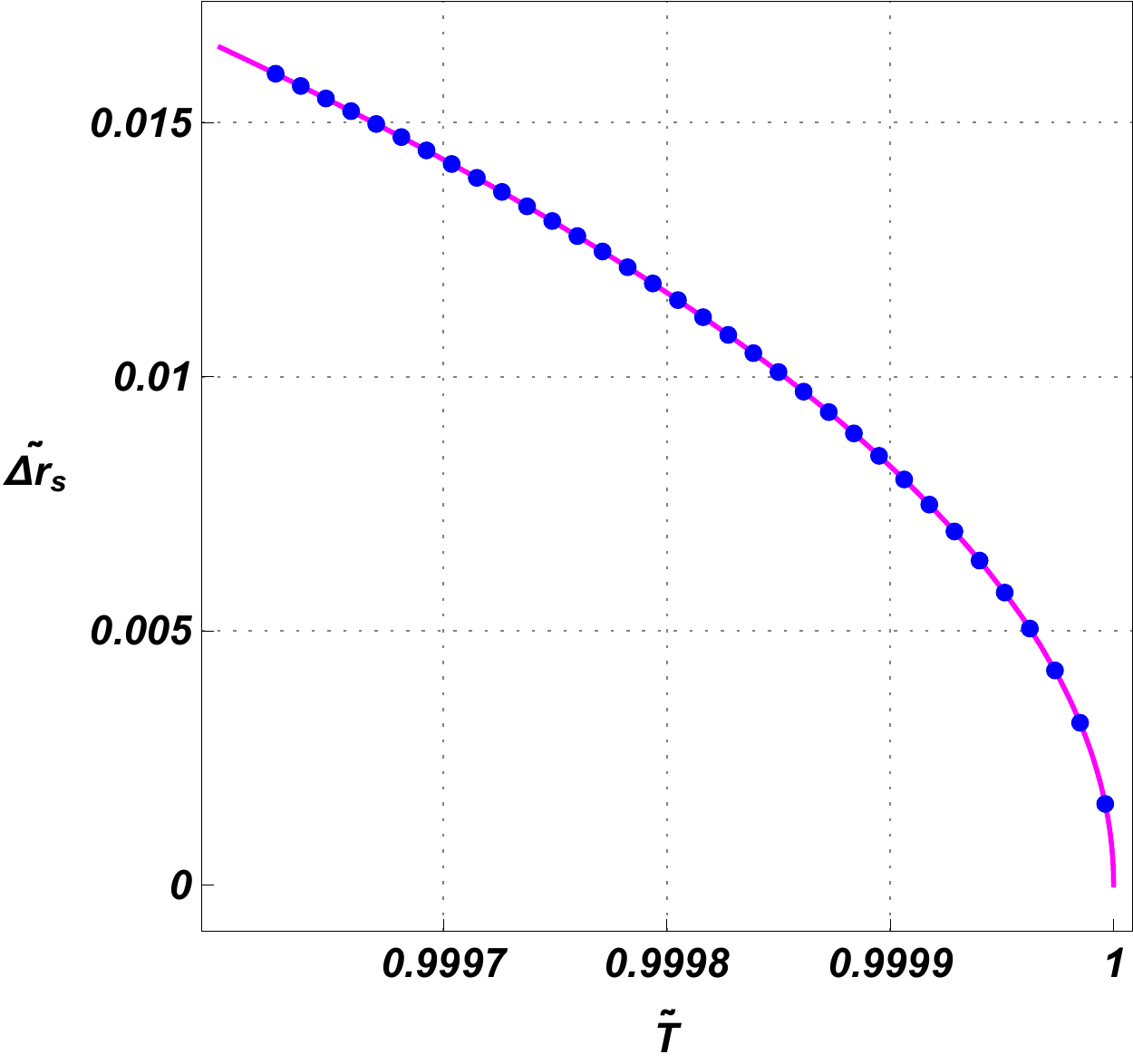}};	
\draw[->,line width=1pt,color=magenta](-4.5,-8.35)--(7.35,2);
\end{tikzpicture}	
\caption{{\it \footnotesize {\bf Left:} The $\Delta \tilde{r}_{s}$ as function of reduced  temperature $\tilde{T}$ The. {\bf Right:} Fitting curves   $\Delta \tilde{r}_{s}$ versus $\tilde{T}_\star$ near the critical temperature. For both panels we have set
$Q=1$.  }}\label{Tdeltaradii}
\end{center}
\end{figure}

In this section, we would like to investigate the phase transition of  four-dimensional  charged   AdS black hole using  shadow formalism.
Considering the temperature expression Eq.\eqref{temp} and the heat capacity Eq.\eqref{CC},   we  can plot the isobar curves on the $T-r_{h,s}$ and $C-r_{h,s}$,  respectively\footnote{The subscript indices $_h$ and $_s$ are associated with the event horizon radius and the black hole shadow radius, respectively.} for a  fixed value of the  charge $Q$. The plots are depicted in Fig.\ref{Tradii}.
They show  obviously  similarities with  the  Van der Waals phase transitions. It follows that,  for different pressure,  such  curves behave differently. For  the large values of pressure $P>P_c$, the temperature involves   monotonic behaviors. While, the   heat capacity is a continuous function  for both variables  $r_h$ and $r_s$.  For small pressures $P<P_c$,  however,  we can observe  that the temperature curves are not monotonic. An examination,   either on   Eq.\eqref{CC} or  on  Fig.\ref{Tradii},  shows   that the specific heat capacity changes its sign. Concretely,  it changes  from positive to negative and then it becomes  positive again.  For an intermediate range of event horizon/shadow radius ($r_h/r_s$),  the black hole is thermodynamically unstable.  In fact, there are three black  holes competing thermodynamically. The smallest black  hole associated with the domain $[0, r_{h_1,s_1}]$ continues to win until the coexistence temperature $T_\star$. Above such a temperature,  the system  is considered as  a  large black hole associated with  the interval $[ r_{h_1,s_2},\infty]$. To remove the unstable intermediate branch associated with the domain $[r_{h_1,s_1},r_{h_2,s_2}]$, we can easily verify the so-called construction of the Maxwell's equal-area interpreted in each plan by
 \begin{equation}\label{maxxx}
\left\{\begin{array}{c} T_\star (r_{h_2}-r_{h_1})=\int_{r_{h_2}}^{r_{h_1}} T dr_h,\\
 T_\star (r_{s_2}-r_{s_1})=\int_{r_{s_2}}^{r_{s_1}} T dr_s\end{array}\right. .
\end{equation}
These two equalities are illustrated in Fig.\ref{Tradii} by the two colored surfaces $\mathcal{A}_1$ and $\mathcal{A}_2$ in each top panel. The Eqs.\eqref{maxxx} are equivalent geometrically to $\mathcal{A}_1=\mathcal{A}_2$,
showing that the  system exhibits  a  phase transition   being  a first-order one.
In the case of    $P=P_c$,    the smallest black hole and the largest one  merge into one squeezing  out the unstable black hole. It has been observed  an inflection point in the $T -r_{h,s}$ planes,  being indicated  by dashed lines   of Fig.\ref{Tradii}. At such a point,  the specific heat capacity is divergent    revealing  that this  phase transition undergoes a second-order phase transition.  In this  way, the critical thermodynamical quantities  can be obtained   by the help of the following equations
\begin{equation}\label{system}
\left.\frac{\partial T}{\partial r_{h,s}}\right|_{P,Q}= \left.\frac{\partial^2 T}{\partial r_{h,s}^2}\right|_{P,Q}=0.
\end{equation}
Substituting Eq.\eqref{temp} into Eq.\eqref{system},  we can find certain critical quantities. Indeed,  the critical pressure, the  critical event horizon radius and the  critical temperature  are given  by
\begin{equation}\label{critic}
P_c=\frac{1}{96\pi Q},\quad r_{h_c}=\sqrt{6}Q\quad \text{ and }\quad T_c= \frac{1}{3\sqrt{6}\pi Q}.
\end{equation}
These give a critical shadow radius which reads  as
\begin{equation}
r_{s_c}=3 \sqrt{\frac{1}{23} \left(29+12 \sqrt{6}\right)} Q\approx 4.78014 Q.
\end{equation}
 It  has been remarked that   the thermodynamical behaviors in $T-r_h$ and  $T-r_s$ planes  share many similarities.
    This could open new gates  to  prob the phase picture of RN-AdS black holes  in terms of  their  shadows. \\
To go beyond such an  investigation, we examine  the behavior of the shadow radius near the second-order phase transition. To see how this works, we evaluate the critical exponent associated with such a  radius.  Indeed, we  first  introduce the following  reduced quantities associated with the temperature, the event horizon radius $r_h$ and the shadow radius $r_s$
\begin{equation}
\tilde{T}=\frac{T}{T_c}, \quad \tilde{r}_h=\frac{r_h}{r_{h c}} \quad \text{and}\quad \tilde{r}_s=\frac{r_s}{r_{s c}}
\end{equation}
which are obtained from  Eq.\eqref{temp}, Eq.\eqref{critic} and Eq.\eqref{rs}. Then, we determine the width of the coexistence lines.  Concretely,   we analyze the behavior of the shadow radius     $\Delta \tilde{r}_{s}$
\begin{equation}\label{delta r0xi0}
  \Delta \tilde{r}_{s}=\tilde{r}_{s_2}-\tilde{r}_{s_1}
  \end{equation}
 before and after the second-order small-large black hole phase transition in $\tilde{T}-\tilde{r}_{s}$ diagram. In the left panel of Fig.\ref{Tdeltaradii}, we  illustrate the difference $\Delta \tilde{r}_{s}$ as a function of the reduced temperature   $\tilde{T}$. It turns out that   this allows one   to  interpret  $\Delta \tilde{r}_{s}$ as an order parameter  corresponding to the phase  transition of    the small-large black holes.  As expected, the width of the coexistence line decreases with the reduced temperature,  having   non zero values at the first-order phase transition. However, it  shrinks  at the critical point.  A close inspection around the second-order phase transition in the right panel of Fig.\ref{Tdeltaradii} shows that the concavity of the graph  is  reversed. In particular,   it  is almost   convex at the first-order phase transition.   However, near  the critical one $T_{c}$,   it becomes  concave.

 To check the consistency of such a finding, we  compute the critical exponent characterizing the universal behavior of the reduced differences $\Delta \tilde{r}_{s}$  around the second-order phase transition. Precisely, we numerically fit the  data according to  the form
  \begin{equation}\label{fit}
 \Delta \tilde{r}_{s} \sim \xi \left(1-\tilde{T}\right)^{\zeta}.
  \end{equation}
 Indeed,  we can  provide  a relation   between  the reduced black hole shadow radius and  the reduced temperature near $T_{c}$. It is given by
  \begin{equation}
  \Delta \tilde{r}_{s}\simeq0.824796 (1 - \tilde{T})^{0.500102}.
  \end{equation}
Under  such a  numerical accuracy, we can obviously claim that the shadow radius around RN-AdS black hole reveals  an  universal critical exponent being close   to $1/2$.  Moreover,  we can cheek  the concavity of the graphs shown in the left panel of Fig.\ref{Tdeltaradii} near  the critical temperature. Using the derivation of  Eq. (\ref{fit})
  \begin{equation}\label{concavity}
  \frac{d^{2}\Delta\tilde{r}_{s}}{d \tilde{T}^{2}}=-\xi \zeta \left(1-\zeta\right)\left(1-\tilde{T}\right)^{\zeta-2}<0,
  \end{equation}
the  graph must be strictly concave when $T$ goes to $ T_{c}$.  We expect that  the  detection of concavity change can be  considered as signatures of  an uncovered  criticality. Therefore,  the black hole shadow radius could  be used  to unveil   the Van der Waals-like phase transition in RN-AdS black hole backgrounds.

  \section{Thermal profile on the shadow silhouette of RN-AdS black hole}\label{sec4}

In this section, we would like to  build  the  so-called thermal image
of such a shadow  black hole in order to  show that
 the complete Van der Waals-like phase picture  reported   in \cite{Kubiznak:2012wp} via  revisited  $T-{r_h}$ curve and  within $T-{r_s}$ diagrams of Fig.\ref{Tradii} can be reflected  in the thermal profile.

\begin{figure}[!ht]
 \center
  \includegraphics[scale=.65]{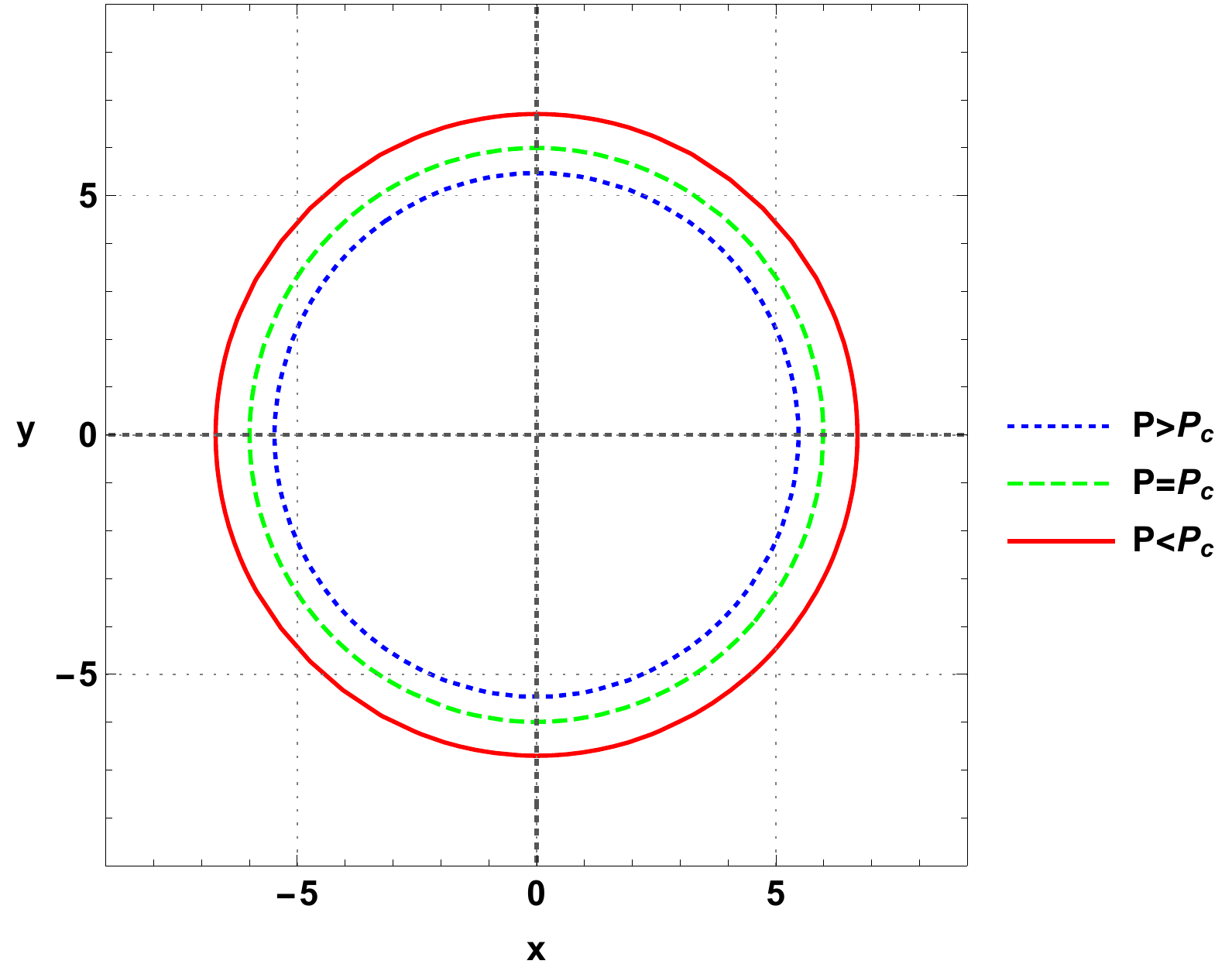}\\
  \caption {\it{ \footnotesize Apparent shape of the charged AdS black hole, as seen by an observer at $(r_o,\alpha= \frac{\pi}{2})$ for different values the the thermodynamical pressure $P$.
 Here we have set the mass $M=60$ and the charge $Q=1$.}}
\label{shadowvsP}
\end{figure}

\begin{figure}[!ht]
		\begin{center}
		\begin{tikzpicture}[scale=0.2,text centered]
\node[] at (-44,1){\small  \includegraphics[scale=0.45]{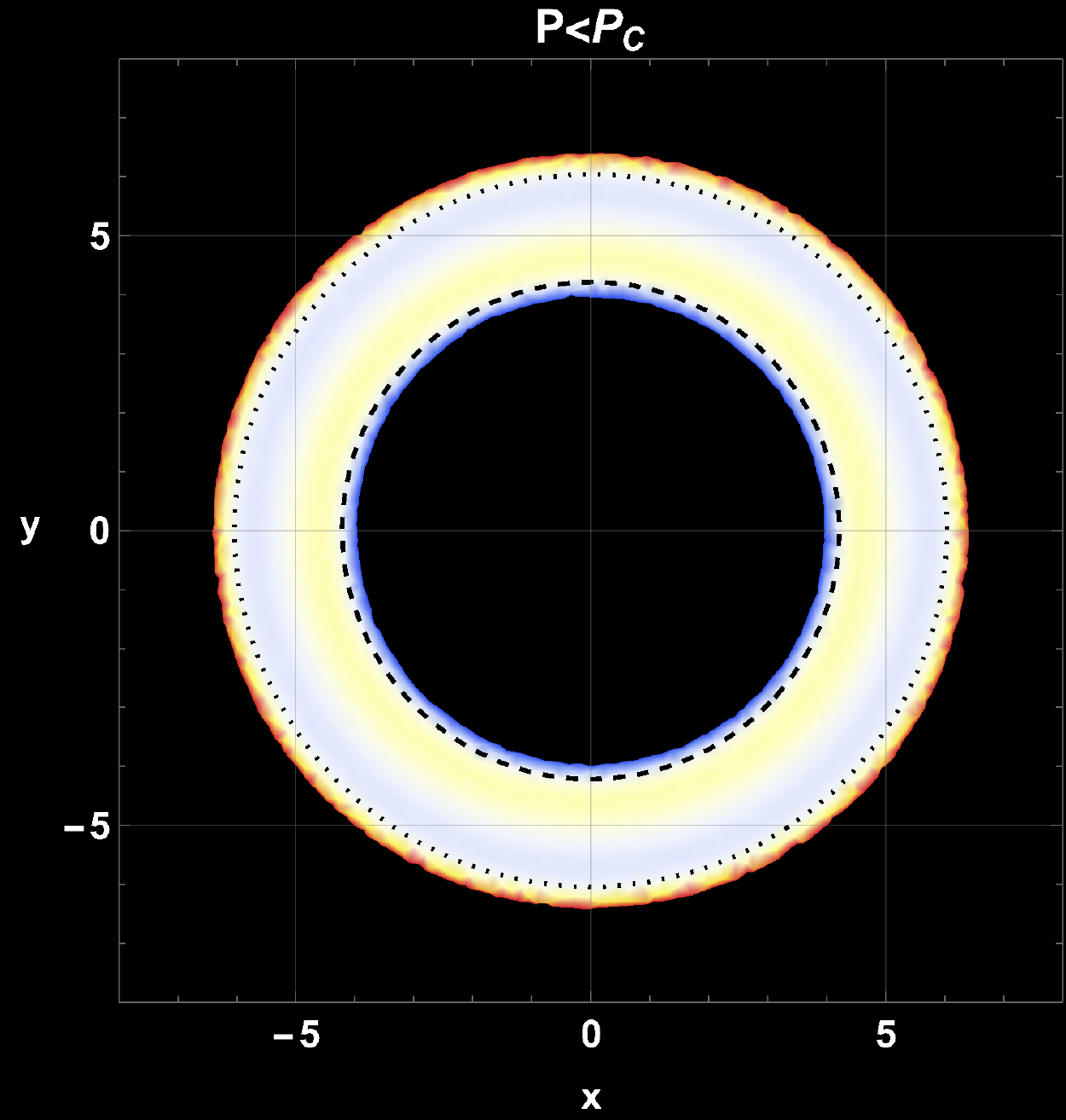}};	


\node[] at (-13,1){\small  \includegraphics[scale=0.45]{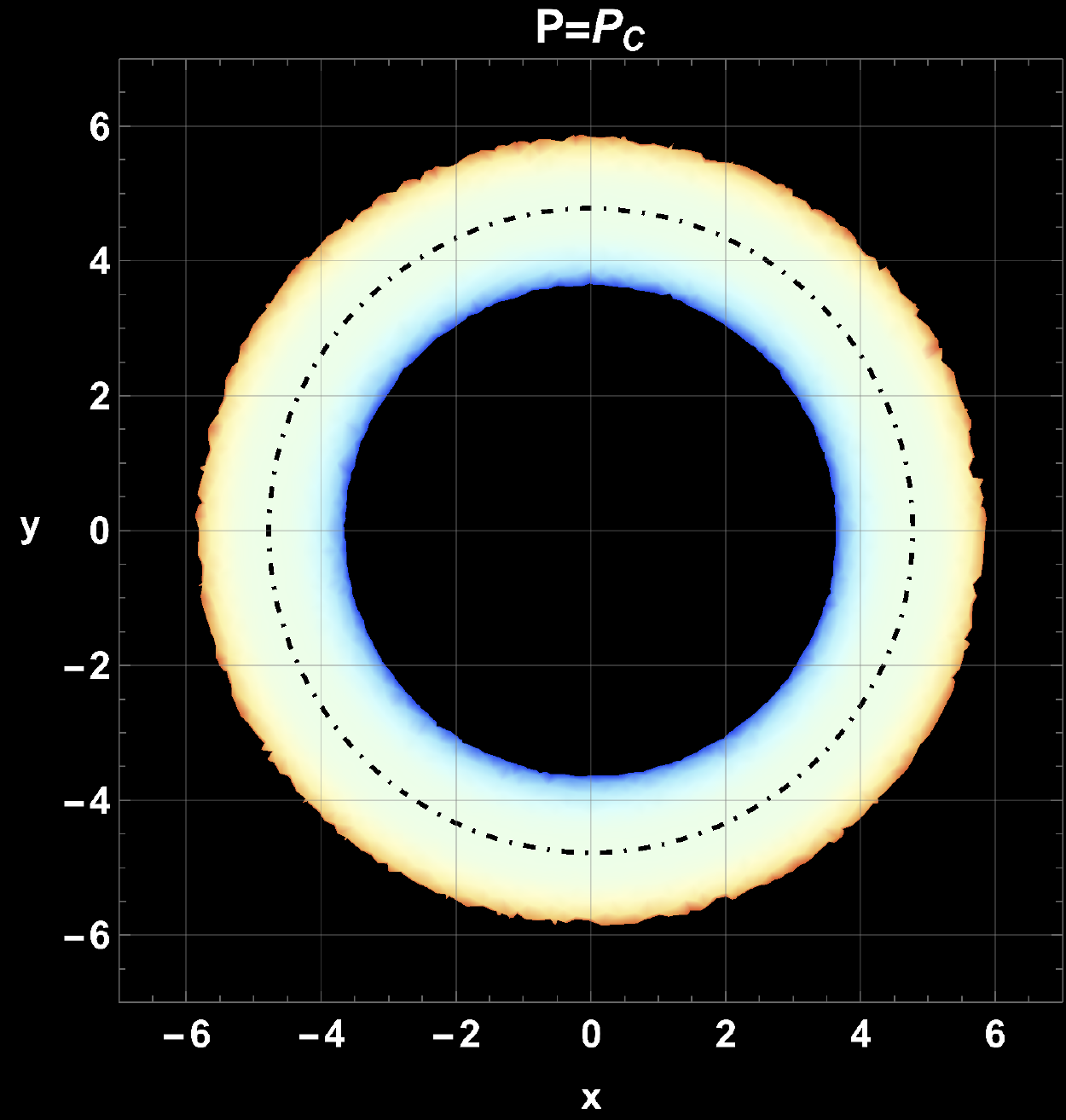}};	
\node[] at (18,1){\small  \includegraphics[scale=0.45]{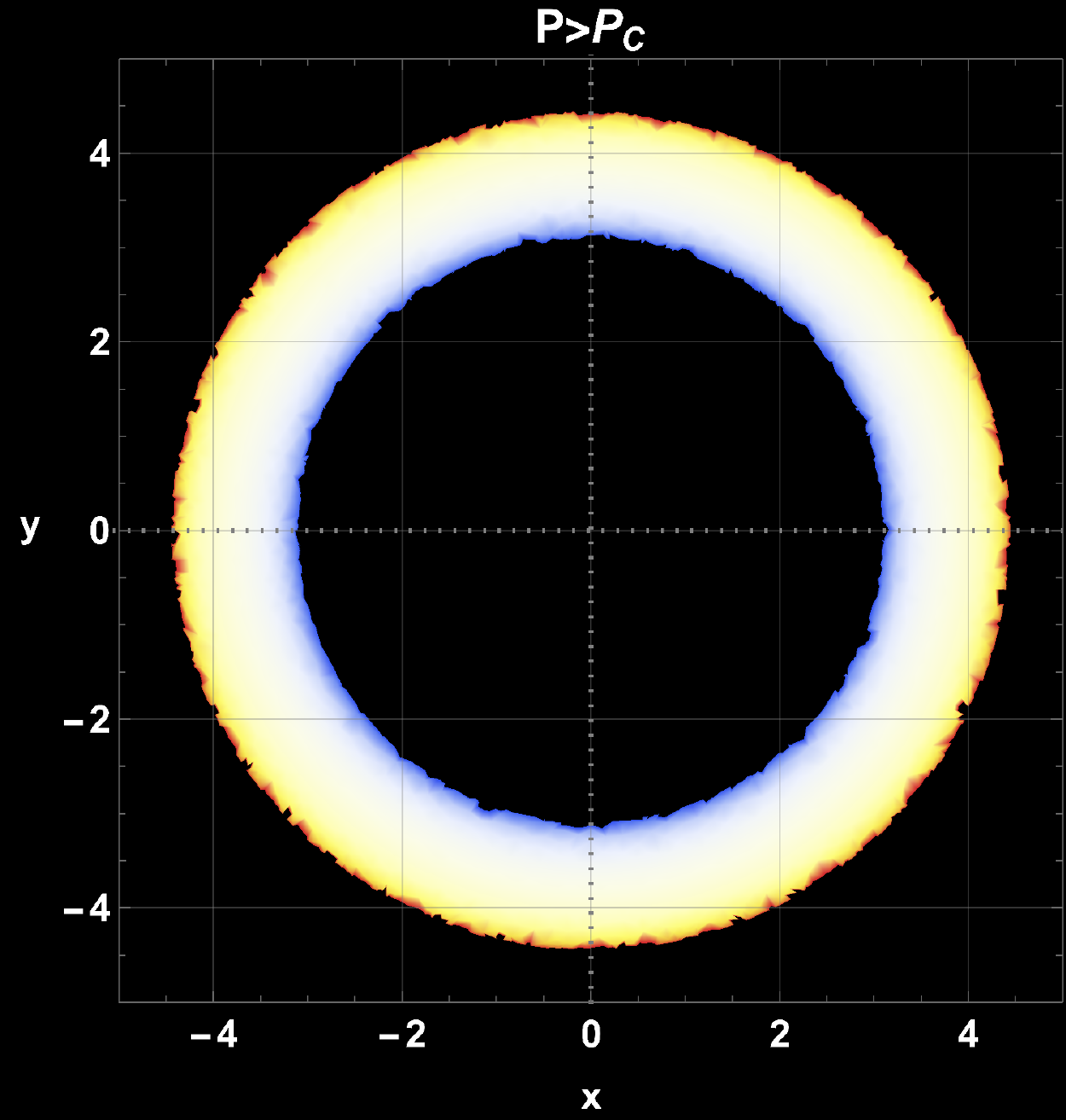}};	
\node[] at (-13,-18.7){\small  \includegraphics[scale=1]{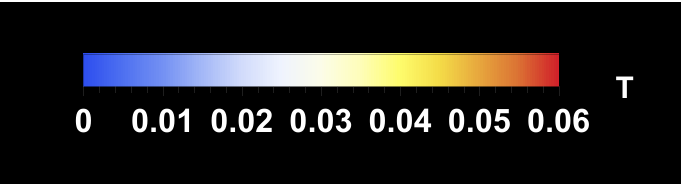}};	
\node[draw ,thin,line width=1pt,color=magenta,name=hete1] at (-28,-38.5){\small  \includegraphics[scale=0.45]{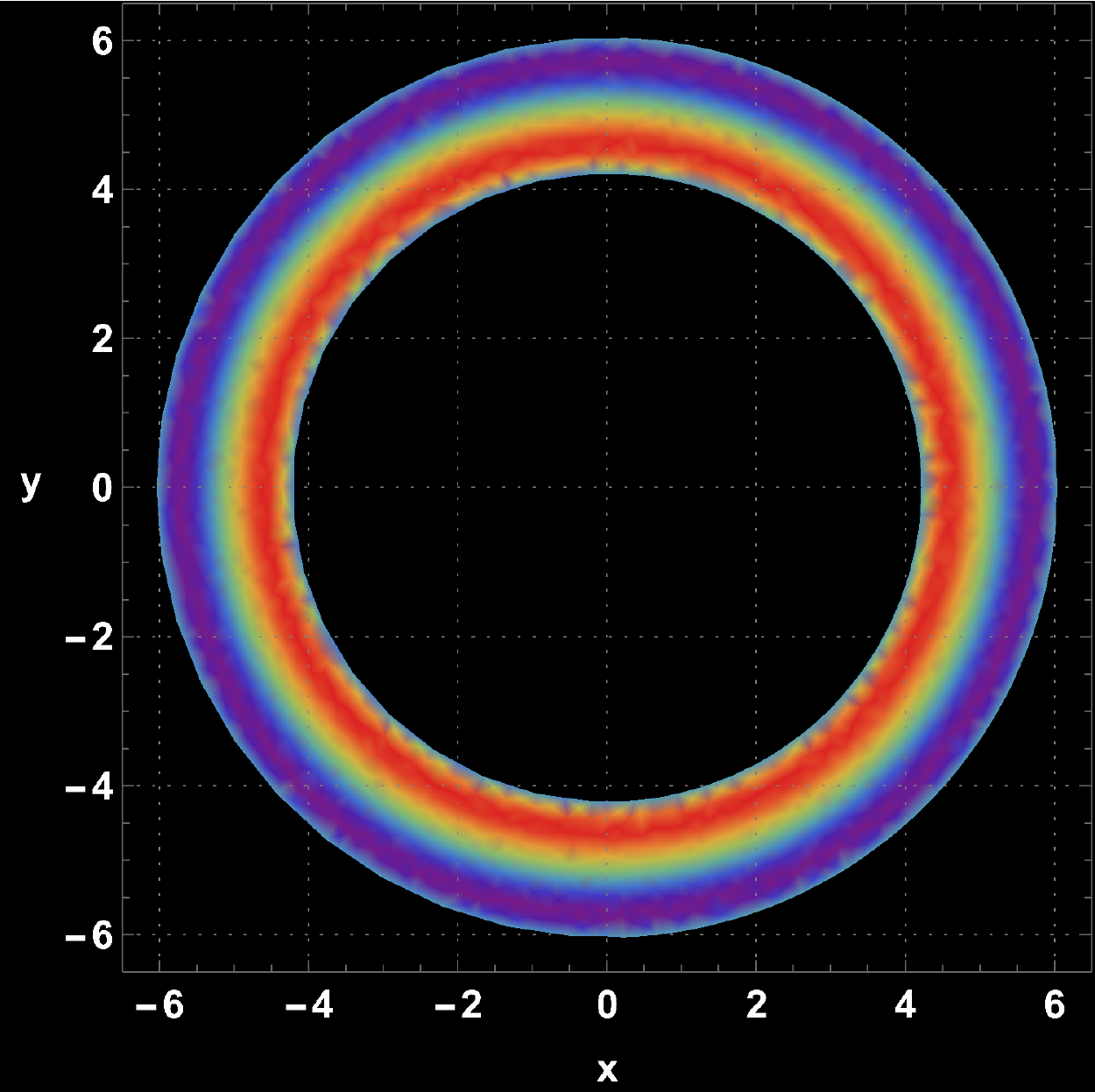}};	
\node[] at (-7.,-38){\small  \includegraphics[scale=0.9]{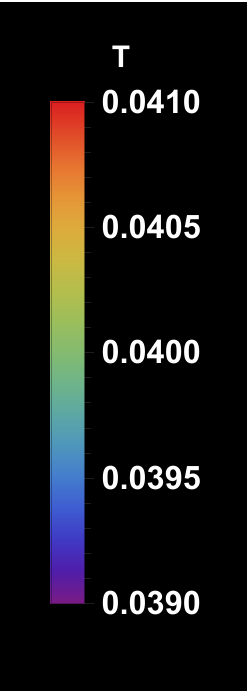}};	
\draw[->,line width=1pt,color=magenta](-39.17,-5.92)--(-32,-23.5);

\draw [magenta,thin,line width=1pt,,decorate,decoration={brace,amplitude=4pt,mirror}
]
(-40,-4.3) -- (-37.5,-6.2);


\node[] at (15,-38){\small  \includegraphics[scale=0.45]{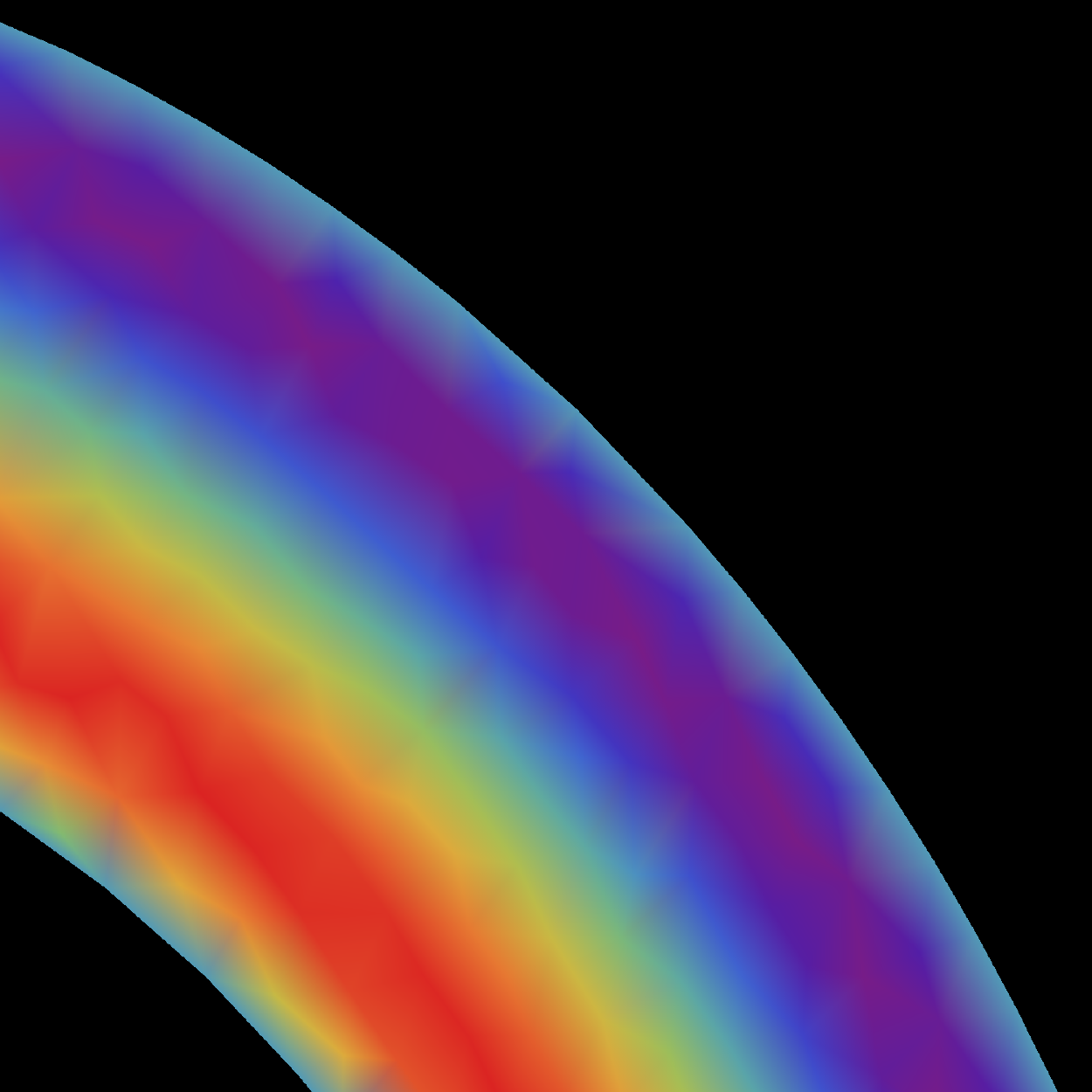}};

\end{tikzpicture}	
\caption{{\it \footnotesize Superposition of the temperature profile on the silhouette of the shadow cast by the RN-AdS black hole for each thermodynamical case. We have set $M=60$, and $Q=1$.}}\label{Tdeltaradiix}
\end{center}
\end{figure}

To visualise  the  shadow boundary curve, we  can exploit  the stereographic projection  in terms of the  cartesian coordinates $(x,y)$,  reported in  \cite{Eiroa:2017uuq},   using  the following equations
\begin{eqnarray}
x&=&\lim_{r\rightarrow\infty}\left(-r^2\sin\theta_0 \frac{d\phi}{dr}\right)_{\theta_0\rightarrow\frac{\pi}{2}}\\
y&=&\lim_{r\rightarrow\infty}\left(r^2 \frac{d\theta}{dr}\right)_{\theta_0\rightarrow\frac{\pi}{2}}.
\end{eqnarray}
Indeed,  we  provide  the behaviour of  the shadow  circular shape  of the RN-AdS black hole  for  different values of the pressure  seen by an observer situated  at $r_o$ and an angle $\alpha$. This is illustrated  in  Fig.\ref{shadowvsP}  for a fixed value of the  charge $Q$. It has been observed   that the size of the circular shape of the black hole shadow depends on the pressure. Concretely,  it becomes  large when $P<P_c$. However, it   decreases as $P$ increase.
To unveil more information on such an  analysis of the black hole phase structure and its shadow, we  implement the temperature  in order to produce  a thermal image. To   do so,   we consider  separately  the shadow silhouette  for the   $P<P_c$, $P=P_c$ and $P>P_c$ cases.  By the use  of existing map between the polar and  the cartesian coordinates,  we turn on   the temperature variation  in terms of $r_s$ on the shadow  circular shape.  This effect  is  plotted  in Fig.\ref{Tdeltaradiix}.

It is noted that the  Fig.\ref{Tdeltaradiix} is  built  from the superposition of each graph from the top-right panel of Fig.\ref{Tradii} and its associated circular shape  presented in Fig.\ref{shadowvsP}. A close examination shows that one has the following findings  corresponding to the above three situations $P<P_c$, $P=P_c$ and $P>P_c$.
  \begin{itemize}
  \item Case $P>P_c$ (Top right panel):\\
    It has been seen that the temperature increase monotonically from the center  of  the  shadow (cold region) to its boundary  (hot region),   which matches perfectly with   Cyan curve of Fig.\ref{Tradii}.
  \item  Case $P=P_c$ (Top middle panel): \\
  The variation is also an increasing  in terms of $r_s$. In this case,  the stagnation near the critical point is interpreted by the large light bluish zone  being  consistent with the dashed green curve appearing  in Fig.\ref{Tradii}.   The dotted-dashed line corresponds to  the critical temperature.
  \item $P<P_c$. (Top left panel)\\
   The global variation of the  temperature in small and large domains is  clearly disclosed within the color gradient of the bar-legend. However,  the  most relevant discussed case is associated with  the temperature. This reveals  an oscillator behavior between the points $r_{s_1}$ and $r_{s_2}$ of the red curve in Fig.\ref{Tradii}  being originated from  the Maxwell construction. This behavior can be also observed from the thermal image where  the radius $r_{s_{1,2}}$ are represented by the dashed and the  dotted circles respectively. At first sight, it  is  not evident to distinguish the oscillation behavior.  To overcome that,  a deeper zoom is realized in the down panels of Fig.\ref{Tdeltaradiix}. Decoding the bar-legend,  it has been remarked   that the variation of the color passes  from
    blue $(T=T_\star=0.0397)$ to red  $(T=0.0410)$ and to darker violet
    $(T=0.0390)$. Then, it  becomes again blue.
   This shows  that the intermediate phase, where the small and large black holes coexist,  is distinctly persisting in the thermal profile.
  \end{itemize}

\begin{figure}[!ht]
		\begin{center}
		\begin{tikzpicture}[scale=0.2,text centered]
\node[] at (-45,1){\small  \includegraphics[scale=0.42]{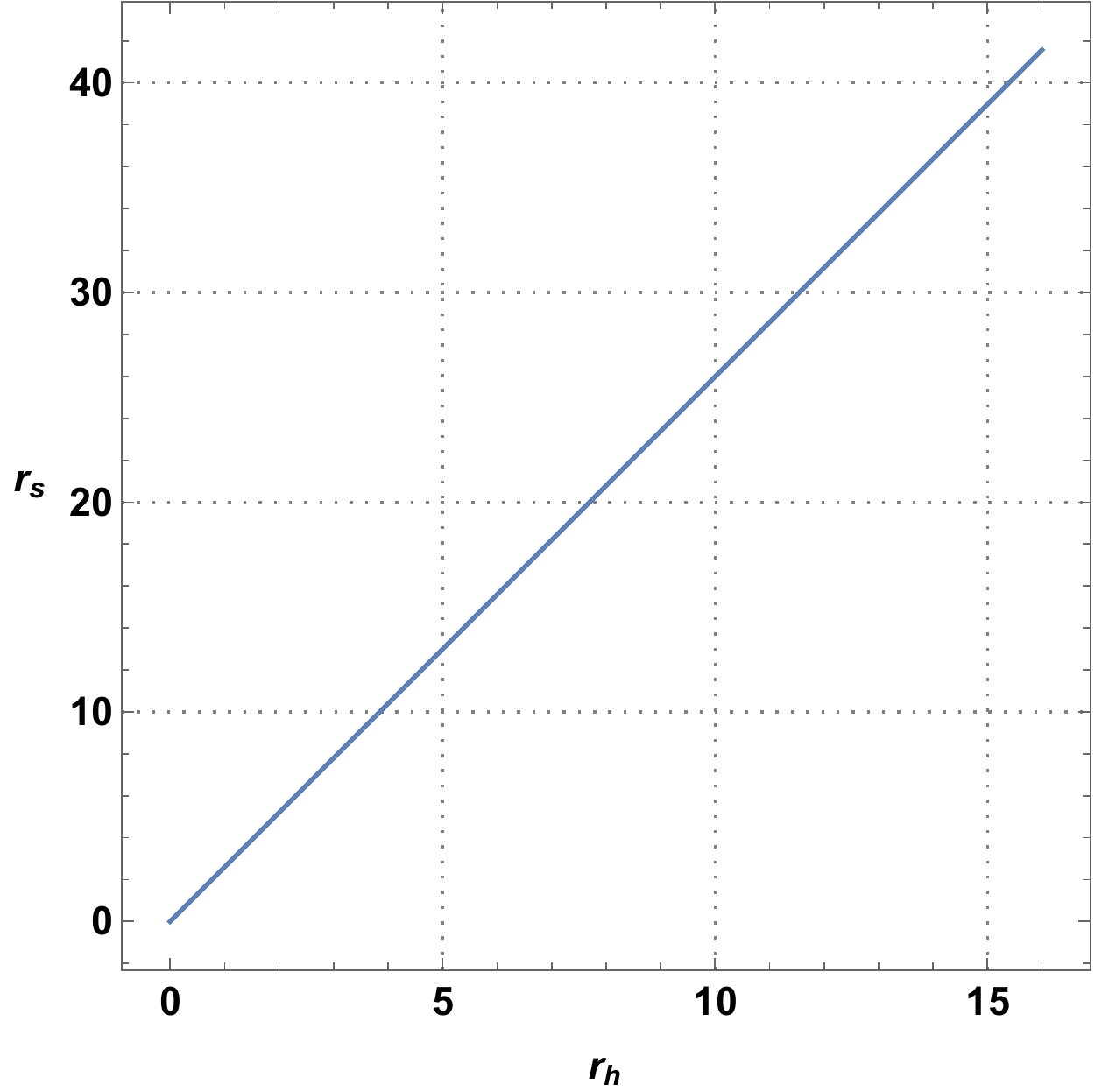}};	


\node[] at (-15.8,2){\small  \includegraphics[scale=0.44]{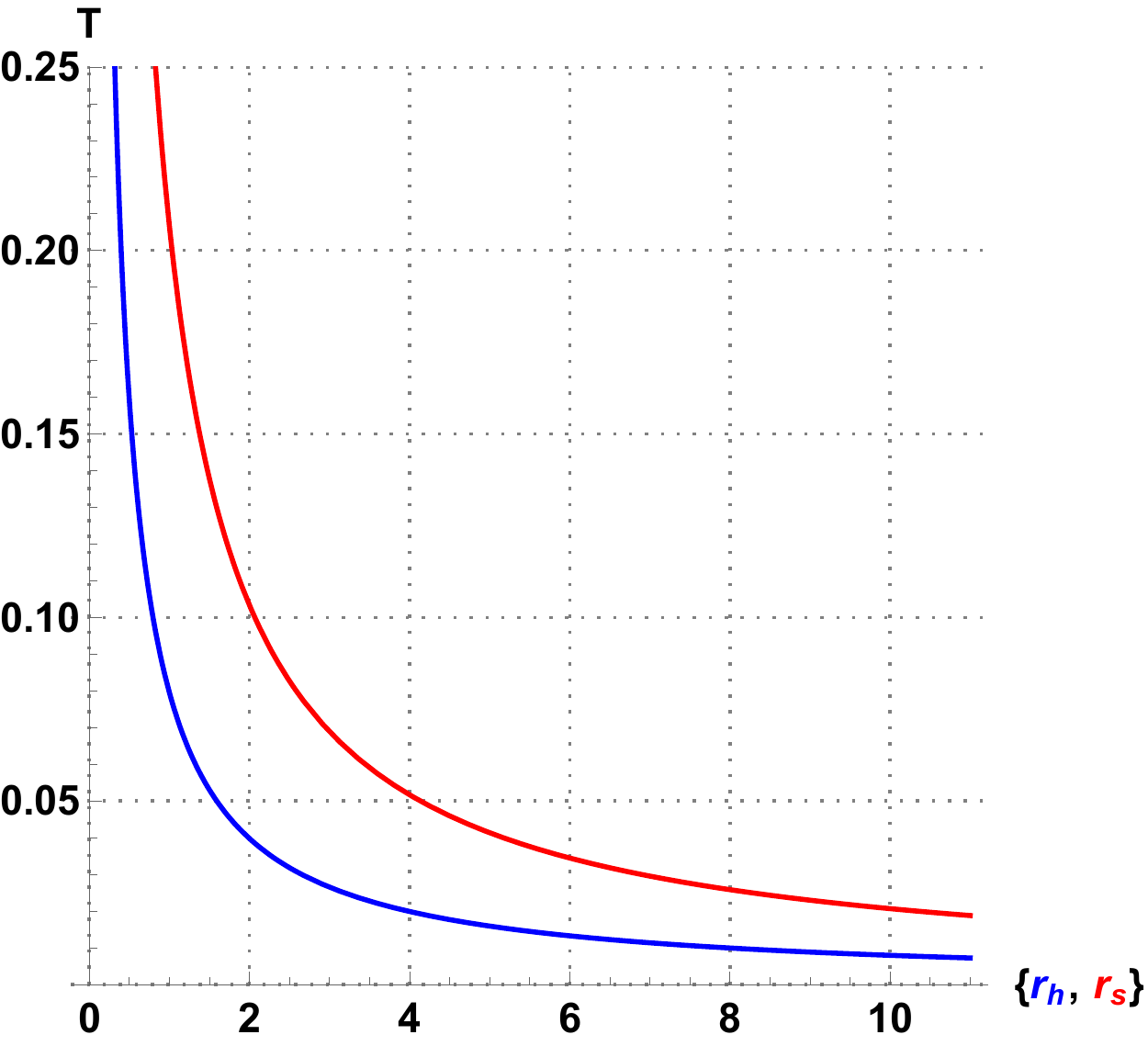}};	
\node[] at (13.3,1){\small  \includegraphics[scale=0.45]{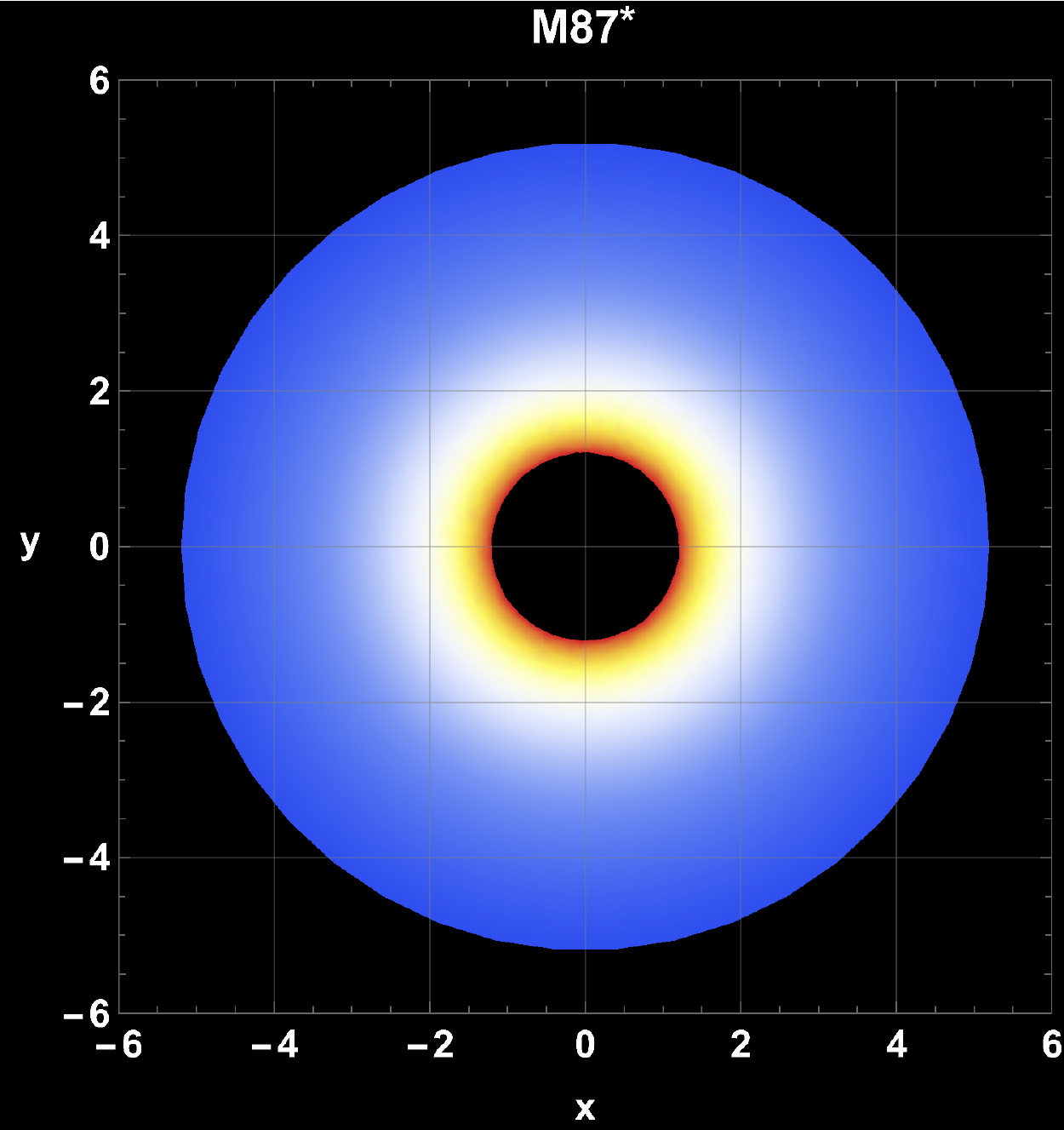}};	
\node[] at (31,-1){\small  \includegraphics[scale=.6]{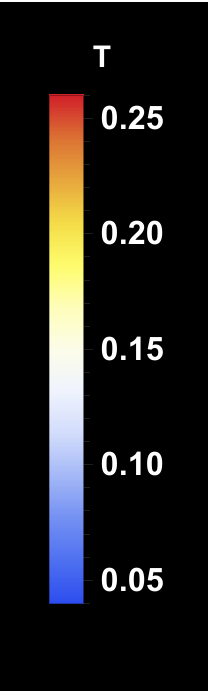}};

\end{tikzpicture}	
\caption{{\it \footnotesize Building  a thermal profile for $M87^\star$ shadow, by Schwarzschild model and  using  $M = 1$ in units of the $M87^\star$ black hole mass given by $M_{BH} =6.5\times 10^{9}M_\odot$ and $r_0 =91.2 kpc$}\cite{Jusufi:2019nrn}.}\label{M87}
\end{center}
\end{figure}

We believe that the obtained results could be seen as a   powerful observational tool to probe the thermodynamics of such a black hole associated with experiment activities including   LIGO, VIRGO and especially the Event horizon telescope.  It is   known that the astrophysical black holes are modeled by the Kerr or  Schwarzschild black hole configurations. In the light of $M87^\star$  observational parameters, we  try to depict a naive version of the thermal profile associated with such a black hole. First, we assume that the $M87^\star$ corresponds to   a Schwarzschild black hole type be setting $Q=P=0$ in the metric equation Eq.\eqref{met1}. Under such considerations, the event horizon, the  Hawking temperature and the shadow radius become respectively  
\begin{equation}
r_h=2M,\qquad T=\frac{1}{4\pi r_h}\qquad \text{and }\qquad r_s=3\sqrt{3}M. 
\end{equation}
The Event Horizon Telescope data  reveals that the  $M87^\star$ black hole involves  a mass $M=(6.5\pm0.7)\times10^9 M_\odot$ and a ring of  a diameter $\theta_d=42\pm3 \mu as$ \cite{Akiyama:2019eap,Akiyama:2019fyp}. With  future improvements of the observations, one can have more key pieces of information about the nature of the black hole space-time.
The same analysis of the previous sections giving  the thermal  profile is   illustrated in  Fig.\ref{M87}

The left panel ensures the validity of this method in the case of Schwarzschild background by checking the condition $\frac{d r_s}{dr_h}>0$. In the middle panel, we have depicted the variation of such a black hole within the event horizon radius $r_h$ and shadow one $r_s$.  we  can easily see that the temperature shares the same decreasing behaviour in terms of $r_{h,s}$.  Lastly, we superpose the variation of temperature on the shadow circular shape to built the thermal profile of the $M87^\star$ under such  assumptions. In this regard, we  can notice that the centre of black hole shadow is hot than its boundary where the temperature approaches zero.

  \section{Conclusion}

  In this  work, we have investigated the shadow of charged AdS black holes being considered as a   new   way to probe  the Van der Waals-like phase picture in the extended phase space. Using  Hamiltonian formalism, we   have, first,  exploited  the equations of motion  controlling  the  photon dynamics  associated with  the equatorial plane in such  curved backgrounds   to build  the associated  shadow  circular shape.
Then,  nice similarities  in the context of  phase transition portrait between  $(T,r_h)$ and  $(T,r_s)$ have been established. For pressure values under the critical one ($P=P_c$), the oscillatory curve of the temperature in terms of $r_s$ as of $r_h$  has provided  the presence of a  coexistence region for small/large black holes showing  that the system undergoes a first-order phase transition.
For  the second-order phase transition, however,  the numerical calculations of the critical exponent in  $\Delta \tilde{r}_{s}-\tilde{T}$ diagram  have generated   a value  being close to  $1/2$  which matches   perfectly  with  the  Van der Waals fluid systems.
  To  support the  present  study,  we  have built  a   thermal image  of such a black hole by varying  the  temperature on the shadow circular  shape.  In  such an image,  we have observed clearly the critical behaviors of RN-AdS black  hole solutions  in the extended phase space.

  This work comes up with certain  open questions.  Among others, it will be interesting to  consider   rotating and dS black hole solutions. Motivated by  EHT image and its related observational data,  it could be  possible to  build a thermal image of real astronomical black holes.  Such  questions might  be addressed  in  future works.
  \\
  \section*{Acknowledgements}
 This work is partially supported by the ICTP through AF-13.

\bibliographystyle{ws-ijmpa}
\bibliography{biblio.bib}

\begin{thebibliography}{10}
\expandafter\ifx\csname urlstyle\endcsname\relax
  \providecommand{\doi}[1]{doi:\discretionary{}{}{}#1}\else
  \providecommand{\doi}{doi:\discretionary{}{}{}\begingroup
  \urlstyle{rm}\Url}\fi

\bibitem{Zhang:2019glo}
M.~Zhang and M.~Guo \href{http://arxiv.org/abs/1909.07033}{{\ttfamily
  arXiv:1909.07033 [gr-qc]}}.

\bibitem{Akiyama:2019cqa}
 Event Horizon Telescope Collaboration (K.~Akiyama {\em et~al.}), {\em
  Astrophys. J.} {\bf 875},  ~L1  (2019),
  \href{http://arxiv.org/abs/1906.11238}{{\ttfamily arXiv:1906.11238
  [astro-ph.GA]}}, \doi{10.3847/2041-8213/ab0ec7}.

\bibitem{Akiyama:2019eap}
 Event Horizon Telescope Collaboration (K.~Akiyama {\em et~al.}), {\em
  Astrophys. J. Lett.} {\bf 875},  ~L6  (2019),
  \href{http://arxiv.org/abs/1906.11243}{{\ttfamily arXiv:1906.11243
  [astro-ph.GA]}}, \doi{10.3847/2041-8213/ab1141}.

\bibitem{Akiyama:2019fyp}
 Event Horizon Telescope Collaboration (K.~Akiyama {\em et~al.}), {\em
  Astrophys. J. Lett.} {\bf 875},  ~L5  (2019),
  \href{http://arxiv.org/abs/1906.11242}{{\ttfamily arXiv:1906.11242
  [astro-ph.GA]}}, \doi{10.3847/2041-8213/ab0f43}.

\bibitem{Abbott:2016blz}
 LIGO Scientific, Virgo Collaboration (B.~Abbott {\em et~al.}), {\em Phys.\
  Rev.\ Lett.} {\bf 116},   061102  (2016),
  \href{http://arxiv.org/abs/1602.03837}{{\ttfamily arXiv:1602.03837 [gr-qc]}},
  \doi{10.1103/PhysRevLett.116.061102}.

\bibitem{Cunha:2019ikd}
P.~V.~P. Cunha, C.~A.~R. Herdeiro and E.~Radu, {\em Universe} {\bf 5},   220
  (2019), \href{http://arxiv.org/abs/1909.08039}{{\ttfamily arXiv:1909.08039
  [gr-qc]}}, \doi{10.3390/universe5120220}.

\bibitem{Vagnozzi:2019apd}
S.~Vagnozzi and L.~Visinelli, {\em Phys. Rev. D} {\bf 100},   024020  (2019),
  \href{http://arxiv.org/abs/1905.12421}{{\ttfamily arXiv:1905.12421 [gr-qc]}},
  \doi{10.1103/PhysRevD.100.024020}.

\bibitem{Allahyari:2019jqz}
A.~Allahyari, M.~Khodadi, S.~Vagnozzi and D.~F. Mota, {\em JCAP} {\bf 02},
  003  (2020), \href{http://arxiv.org/abs/1912.08231}{{\ttfamily
  arXiv:1912.08231 [gr-qc]}}, \doi{10.1088/1475-7516/2020/02/003}.

\bibitem{synge1966escape}
J.~Synge, {\em Monthly Notices of the Royal Astronomical Society} {\bf 131},
  463  (1966).

\bibitem{luminet1979image}
J.-P. Luminet, {\em Astronomy and Astrophysics} {\bf 75}, 228  (1979).

\bibitem{hawking1973black}
S.~Hawking, B.~Carter, J.~M. Bardeen, H.~Gursky, K.~S. Thorne, R.~Ruffini,
  I.~D. Novikov {\em et~al.}, {\em Black holes} (CRC Press, 1973).

\bibitem{de2000apparent}
A.~De~Vries, {\em Classical and Quantum Gravity} {\bf 17},   123  (2000).

\bibitem{Hioki:2008zw}
K.~Hioki and U.~Miyamoto, {\em Phys. Rev.} {\bf D78},   044007  (2008),
  \href{http://arxiv.org/abs/0805.3146}{{\ttfamily arXiv:0805.3146 [gr-qc]}},
  \doi{10.1103/PhysRevD.78.044007}.

\bibitem{Li:2020drn}
P.-C. Li, M.~Guo and B.~Chen, {\em Phys. Rev. D} {\bf 101},   084041  (2020),
  \href{http://arxiv.org/abs/2001.04231}{{\ttfamily arXiv:2001.04231 [gr-qc]}},
  \doi{10.1103/PhysRevD.101.084041}.

\bibitem{Guo:2019lur}
M.~Guo, S.~Song and H.~Yan, {\em Phys. Rev. D} {\bf 101},   024055  (2020),
  \href{http://arxiv.org/abs/1911.04796}{{\ttfamily arXiv:1911.04796 [gr-qc]}},
  \doi{10.1103/PhysRevD.101.024055}.

\bibitem{Shaikh:2018lcc}
R.~Shaikh, P.~Kocherlakota, R.~Narayan and P.~S. Joshi, {\em Mon. Not. Roy.
  Astron. Soc.} {\bf 482}, 52  (2019),
  \href{http://arxiv.org/abs/1802.08060}{{\ttfamily arXiv:1802.08060
  [astro-ph.HE]}}, \doi{10.1093/mnras/sty2624}.

\bibitem{Grenzebach:2014fha}
A.~Grenzebach, V.~Perlick and C.~L{\"a}mmerzahl, {\em Phys. Rev.} {\bf D89},
  124004  (2014), \href{http://arxiv.org/abs/1403.5234}{{\ttfamily
  arXiv:1403.5234 [gr-qc]}}, \doi{10.1103/PhysRevD.89.124004}.

\bibitem{Haroon:2018ryd}
S.~Haroon, M.~Jamil, K.~Jusufi, K.~Lin and R.~B. Mann, {\em Phys. Rev.} {\bf
  D99},   044015  (2019), \href{http://arxiv.org/abs/1810.04103}{{\ttfamily
  arXiv:1810.04103 [gr-qc]}}, \doi{10.1103/PhysRevD.99.044015}.

\bibitem{Johannsen:2015hib}
T.~Johannsen, A.~E. Broderick, P.~M. Plewa, S.~Chatzopoulos, S.~S. Doeleman,
  F.~Eisenhauer, V.~L. Fish, R.~Genzel, O.~Gerhard and M.~D. Johnson, {\em
  Phys. Rev. Lett.} {\bf 116},   031101  (2016),
  \href{http://arxiv.org/abs/1512.02640}{{\ttfamily arXiv:1512.02640
  [astro-ph.GA]}}, \doi{10.1103/PhysRevLett.116.031101}.

\bibitem{Cunha:2015yba}
P.~V.~P. Cunha, C.~A.~R. Herdeiro, E.~Radu and H.~F. Runarsson, {\em Phys. Rev.
  Lett.} {\bf 115},   211102  (2015),
  \href{http://arxiv.org/abs/1509.00021}{{\ttfamily arXiv:1509.00021 [gr-qc]}},
  \doi{10.1103/PhysRevLett.115.211102}.

\bibitem{Eiroa:2017uuq}
E.~F. Eiroa and C.~M. Sendra, {\em Eur. Phys. J.} {\bf C78},  ~91  (2018),
  \href{http://arxiv.org/abs/1711.08380}{{\ttfamily arXiv:1711.08380 [gr-qc]}},
  \doi{10.1140/epjc/s10052-018-5586-6}.

\bibitem{Wang:2017qhh}
M.~Wang, S.~Chen and J.~Jing, {\em Phys. Rev.} {\bf D97},   064029  (2018),
  \href{http://arxiv.org/abs/1710.07172}{{\ttfamily arXiv:1710.07172 [gr-qc]}},
  \doi{10.1103/PhysRevD.97.064029}.

\bibitem{Tsukamoto:2017fxq}
N.~Tsukamoto, {\em Phys. Rev.} {\bf D97},   064021  (2018),
  \href{http://arxiv.org/abs/1708.07427}{{\ttfamily arXiv:1708.07427 [gr-qc]}},
  \doi{10.1103/PhysRevD.97.064021}.

\bibitem{Tsupko:2017rdo}
O.~{\relax Yu}. Tsupko, {\em Phys. Rev.} {\bf D95},   104058  (2017),
  \href{http://arxiv.org/abs/1702.04005}{{\ttfamily arXiv:1702.04005 [gr-qc]}},
  \doi{10.1103/PhysRevD.95.104058}.

\bibitem{Sharif:2016znp}
M.~Sharif and S.~Iftikhar, {\em Eur. Phys. J.} {\bf C76},   630  (2016),
  \href{http://arxiv.org/abs/1611.00611}{{\ttfamily arXiv:1611.00611 [gr-qc]}},
  \doi{10.1140/epjc/s10052-016-4472-3}.

\bibitem{Ohgami:2016iqm}
T.~Ohgami and N.~Sakai, {\em Phys. Rev.} {\bf D94},   064071  (2016),
  \href{http://arxiv.org/abs/1704.07093}{{\ttfamily arXiv:1704.07093 [gr-qc]}},
  \doi{10.1103/PhysRevD.94.064071}.

\bibitem{Younsi:2016azx}
Z.~Younsi, A.~Zhidenko, L.~Rezzolla, R.~Konoplya and Y.~Mizuno, {\em Phys.
  Rev.} {\bf D94},   084025  (2016),
  \href{http://arxiv.org/abs/1607.05767}{{\ttfamily arXiv:1607.05767 [gr-qc]}},
  \doi{10.1103/PhysRevD.94.084025}.

\bibitem{Abdujabbarov:2016hnw}
A.~Abdujabbarov, M.~Amir, B.~Ahmedov and S.~G. Ghosh, {\em Phys. Rev.} {\bf
  D93},   104004  (2016), \href{http://arxiv.org/abs/1604.03809}{{\ttfamily
  arXiv:1604.03809 [gr-qc]}}, \doi{10.1103/PhysRevD.93.104004}.

\bibitem{Amir:2016cen}
M.~Amir and S.~G. Ghosh, {\em Phys. Rev.} {\bf D94},   024054  (2016),
  \href{http://arxiv.org/abs/1603.06382}{{\ttfamily arXiv:1603.06382 [gr-qc]}},
  \doi{10.1103/PhysRevD.94.024054}.

\bibitem{Atamurotov:2015nra}
F.~Atamurotov and B.~Ahmedov, {\em Phys. Rev.} {\bf D92},   084005  (2015),
  \href{http://arxiv.org/abs/1507.08131}{{\ttfamily arXiv:1507.08131 [gr-qc]}},
  \doi{10.1103/PhysRevD.92.084005}.

\bibitem{Perlick:2015vta}
V.~Perlick, O.~{\relax Yu}. Tsupko and G.~S. Bisnovatyi-Kogan, {\em Phys. Rev.}
  {\bf D92},   104031  (2015),
  \href{http://arxiv.org/abs/1507.04217}{{\ttfamily arXiv:1507.04217 [gr-qc]}},
  \doi{10.1103/PhysRevD.92.104031}.

\bibitem{Moffat:2015kva}
J.~W. Moffat, {\em Eur. Phys. J.} {\bf C75},   130  (2015),
  \href{http://arxiv.org/abs/1502.01677}{{\ttfamily arXiv:1502.01677 [gr-qc]}},
  \doi{10.1140/epjc/s10052-015-3352-6}.

\bibitem{Lu:2014zja}
R.-S. Lu, A.~E. Broderick, F.~Baron, J.~D. Monnier, V.~L. Fish, S.~S. Doeleman
  and V.~Pankratius, {\em Astrophys. J.} {\bf 788},   120  (2014),
  \href{http://arxiv.org/abs/1404.7095}{{\ttfamily arXiv:1404.7095
  [astro-ph.IM]}}, \doi{10.1088/0004-637X/788/2/120}.

\bibitem{Atamurotov:2013sca}
F.~Atamurotov, A.~Abdujabbarov and B.~Ahmedov, {\em Phys. Rev.} {\bf D88},
  064004  (2013), \doi{10.1103/PhysRevD.88.064004}.

\bibitem{Guo:2018kis}
M.~Guo, N.~A. Obers and H.~Yan, {\em Phys. Rev.} {\bf D98},   084063  (2018),
  \href{http://arxiv.org/abs/1806.05249}{{\ttfamily arXiv:1806.05249 [gr-qc]}},
  \doi{10.1103/PhysRevD.98.084063}.

\bibitem{Yan:2019etp}
H.~Yan, {\em Phys. Rev.} {\bf D99},   084050  (2019),
  \href{http://arxiv.org/abs/1903.04382}{{\ttfamily arXiv:1903.04382 [gr-qc]}},
  \doi{10.1103/PhysRevD.99.084050}.

\bibitem{Hennigar:2018hza}
R.~A. Hennigar, M.~B.~J. Poshteh and R.~B. Mann, {\em Phys. Rev.} {\bf D97},
  064041  (2018), \href{http://arxiv.org/abs/1801.03223}{{\ttfamily
  arXiv:1801.03223 [gr-qc]}}, \doi{10.1103/PhysRevD.97.064041}.

\bibitem{Konoplya:2019sns}
R.~A. Konoplya, {\em Phys. Lett.} {\bf B795}, 1  (2019),
  \href{http://arxiv.org/abs/1905.00064}{{\ttfamily arXiv:1905.00064 [gr-qc]}},
  \doi{10.1016/j.physletb.2019.05.043}.

\bibitem{Bambi:2008jg}
C.~Bambi and K.~Freese, {\em Phys. Rev.} {\bf D79},   043002  (2009),
  \href{http://arxiv.org/abs/0812.1328}{{\ttfamily arXiv:0812.1328
  [astro-ph]}}, \doi{10.1103/PhysRevD.79.043002}.

\bibitem{Bambi:2010hf}
C.~Bambi and N.~Yoshida, {\em Class. Quant. Grav.} {\bf 27},   205006  (2010),
  \href{http://arxiv.org/abs/1004.3149}{{\ttfamily arXiv:1004.3149 [gr-qc]}},
  \doi{10.1088/0264-9381/27/20/205006}.

\bibitem{Konoplya:2019fpy}
R.~A. Konoplya, T.~Pappas and A.~Zhidenko, {\em Phys. Rev. D} {\bf 101},
  044054  (2020), \href{http://arxiv.org/abs/1907.10112}{{\ttfamily
  arXiv:1907.10112 [gr-qc]}}, \doi{10.1103/PhysRevD.101.044054}.

\bibitem{Bambi:2019tjh}
C.~Bambi, K.~Freese, S.~Vagnozzi and L.~Visinelli, {\em Phys. Rev. D} {\bf
  100},   044057  (2019), \href{http://arxiv.org/abs/1904.12983}{{\ttfamily
  arXiv:1904.12983 [gr-qc]}}, \doi{10.1103/PhysRevD.100.044057}.

\bibitem{Chamblin:1999tk}
A.~Chamblin, R.~Emparan, C.~V. Johnson and R.~C. Myers, {\em Phys. Rev.} {\bf
  D60},   064018  (1999), \href{http://arxiv.org/abs/hep-th/9902170}{{\ttfamily
  arXiv:hep-th/9902170 [hep-th]}}, \doi{10.1103/PhysRevD.60.064018}.

\bibitem{Kubiznak:2012wp}
D.~Kubiznak and R.~B. Mann, {\em JHEP} {\bf 07},   033  (2012),
  \href{http://arxiv.org/abs/1205.0559}{{\ttfamily arXiv:1205.0559 [hep-th]}},
  \doi{10.1007/JHEP07(2012)033}.

\bibitem{Belhaj:2012bg}
A.~Belhaj, M.~Chabab, H.~El~Moumni and M.~B. Sedra, {\em Chin. Phys. Lett.}
  {\bf 29},   100401  (2012), \href{http://arxiv.org/abs/1210.4617}{{\ttfamily
  arXiv:1210.4617 [hep-th]}}, \doi{10.1088/0256-307X/29/10/100401}.

\bibitem{Gunasekaran:2012dq}
S.~Gunasekaran, R.~B. Mann and D.~Kubiznak, {\em JHEP} {\bf 11},   110  (2012),
  \href{http://arxiv.org/abs/1208.6251}{{\ttfamily arXiv:1208.6251 [hep-th]}},
  \doi{10.1007/JHEP11(2012)110}.

\bibitem{Belhaj:2015hha}
A.~Belhaj, M.~Chabab, H.~El~Moumni, K.~Masmar, M.~B. Sedra and A.~Segui, {\em
  JHEP} {\bf 05},   149  (2015),
  \href{http://arxiv.org/abs/1503.07308}{{\ttfamily arXiv:1503.07308
  [hep-th]}}, \doi{10.1007/JHEP05(2015)149}.

\bibitem{Hendi:2012um}
S.~H. Hendi and M.~H. Vahidinia, {\em Phys. Rev.} {\bf D88},   084045  (2013),
  \href{http://arxiv.org/abs/1212.6128}{{\ttfamily arXiv:1212.6128 [hep-th]}},
  \doi{10.1103/PhysRevD.88.084045}.

\bibitem{Zhang:2015ova}
J.-L. Zhang, R.-G. Cai and H.~Yu, {\em Phys. Rev.} {\bf D91},   044028  (2015),
  \href{http://arxiv.org/abs/1502.01428}{{\ttfamily arXiv:1502.01428
  [hep-th]}}, \doi{10.1103/PhysRevD.91.044028}.

\bibitem{Wei:2015iwa}
S.-W. Wei and Y.-X. Liu, {\em Phys. Rev. Lett.} {\bf 115},   111302  (2015),
  \href{http://arxiv.org/abs/1502.00386}{{\ttfamily arXiv:1502.00386 [gr-qc]}},
  \doi{10.1103/PhysRevLett.116.169903, 10.1103/PhysRevLett.115.111302},
  [Erratum: Phys. Rev. Lett.116,no.16,169903(2016)].

\bibitem{Chabab:2015ytz}
M.~Chabab, H.~El~Moumni and K.~Masmar, {\em Eur. Phys. J.} {\bf C76},   304
  (2016), \href{http://arxiv.org/abs/1512.07832}{{\ttfamily arXiv:1512.07832
  [hep-th]}}, \doi{10.1140/epjc/s10052-016-4155-0}.

\bibitem{Perlick:2018iye}
V.~Perlick, O.~Y. Tsupko and G.~S. Bisnovatyi-Kogan, {\em Phys. Rev. D} {\bf
  97},   104062  (2018), \href{http://arxiv.org/abs/1804.04898}{{\ttfamily
  arXiv:1804.04898 [gr-qc]}}, \doi{10.1103/PhysRevD.97.104062}.

\bibitem{Nguyen:2015wfa}
P.~H. Nguyen, {\em JHEP} {\bf 12},   139  (2015),
  \href{http://arxiv.org/abs/1508.01955}{{\ttfamily arXiv:1508.01955
  [hep-th]}}, \doi{10.1007/JHEP12(2015)139}.

\bibitem{moiplb}
H.~El~Moumni, {\em Phys. Lett.} {\bf B776}, 124  (2018),
  \doi{10.1016/j.physletb.2017.11.037}.

\bibitem{Liu:2014gvf}
Y.~Liu, D.-C. Zou and B.~Wang, {\em JHEP} {\bf 09},   179  (2014),
  \href{http://arxiv.org/abs/1405.2644}{{\ttfamily arXiv:1405.2644 [hep-th]}},
  \doi{10.1007/JHEP09(2014)179}.

\bibitem{Chabab:2016cem}
M.~Chabab, H.~El~Moumni, S.~Iraoui and K.~Masmar, {\em Eur. Phys. J.} {\bf
  C76},   676  (2016), \href{http://arxiv.org/abs/1606.08524}{{\ttfamily
  arXiv:1606.08524 [hep-th]}}, \doi{10.1140/epjc/s10052-016-4518-6}.

\bibitem{Zou:2017juz}
D.-C. Zou, Y.~Liu and R.-H. Yue, {\em Eur. Phys. J.} {\bf C77},   365  (2017),
  \href{http://arxiv.org/abs/1702.08118}{{\ttfamily arXiv:1702.08118 [gr-qc]}},
  \doi{10.1140/epjc/s10052-017-4937-z}.

\bibitem{Chabab:2018lzf}
M.~Chabab, H.~El~Moumni, S.~Iraoui, K.~Masmar and S.~Zhizeh, {\em Phys. Lett.}
  {\bf B781}, 316  (2018), \href{http://arxiv.org/abs/1804.03960}{{\ttfamily
  arXiv:1804.03960 [hep-th]}}, \doi{10.1016/j.physletb.2018.04.014}.

\bibitem{Wei:2017mwc}
S.-W. Wei and Y.-X. Liu, {\em Phys. Rev.} {\bf D97},   104027  (2018),
  \href{http://arxiv.org/abs/1711.01522}{{\ttfamily arXiv:1711.01522 [gr-qc]}},
  \doi{10.1103/PhysRevD.97.104027}.

\bibitem{Chabab:2019kfs}
M.~Chabab, H.~El~Moumni, S.~Iraoui and K.~Masmar, {\em Int. J. Mod. Phys.} {\bf
  A34},   1950231  (2020), \href{http://arxiv.org/abs/1902.00557}{{\ttfamily
  arXiv:1902.00557 [hep-th]}}, \doi{10.1142/S0217751X19502312}.

\bibitem{Jusufi:2019nrn}
K.~Jusufi, M.~Jamil, P.~Salucci, T.~Zhu and S.~Haroon, {\em Phys. Rev. D} {\bf
  100},   044012  (2019), \href{http://arxiv.org/abs/1905.11803}{{\ttfamily
  arXiv:1905.11803 [physics.gen-ph]}}, \doi{10.1103/PhysRevD.100.044012}.

\end{thebibliography}

\end{document}